\documentclass[aps,prb,twocolumn,superscriptaddress,longbibliography]{revtex4-2}

\usepackage[caption=false]{subfig}
\usepackage{amsmath}
\usepackage{amssymb}
\usepackage{graphicx}
\usepackage{dcolumn}
\usepackage{bm}
\usepackage{textcomp}
\usepackage{braket}
\usepackage{comment}
\usepackage{siunitx}
\usepackage{upgreek}
\usepackage{mathrsfs}
\usepackage{multirow}
\usepackage{physics}
\usepackage{siunitx}
\usepackage{natbib}
\usepackage[hypertexnames=false]{hyperref}

\usepackage{booktabs}
\usepackage{wasysym}
\usepackage{comment}

\def\hto{$\text{Ho}_2\text{Ti}_2\text{O}_7$}

\def\yto{$\text{Yb}_2\text{Ti}_2\text{O}_7$}

\def\cso{$\text{Ce}_2\text{Sn}_2\text{O}_7$}
\def\cho{$\text{Ce}_2\text{Hf}_2\text{O}_7$}
\def\nzo{$\text{Nd}_2\text{Zr}_2\text{O}_7$}
\def\nso{$\text{Nd}_2\text{Sn}_2\text{O}_7$}
\def\nho{$\text{Nd}_2\text{Hf}_2\text{O}_7$}

\def\uB{\mathrm{\upmu_B}}

\begin{document}

\title{Dynamical moment fragmentation in the all-in all-out pyrochlore \texorpdfstring{Nd$_2$Sn$_2$O$_7$}{Nd2Sn2O7}}

\author{Yi~Luo}
\email{luoy1@ornl.gov}
\affiliation{
	Neutron Scattering Division, Oak Ridge National Laboratory, Oak Ridge, Tennessee 37831, USA}

\author{Matthew~S.~Powell}
\affiliation{Department of Chemistry, Clemson University, Clemson, South Carolina 29634-0973, USA}

\author{Joseph~A.~M.~Paddison}
\affiliation{
	Neutron Scattering Division, Oak Ridge National Laboratory, Oak Ridge, Tennessee 37831, USA}
      
\author{Brenden~R.~Ortiz}
\affiliation{Materials Science and Technology Division, Oak Ridge National Laboratory, Oak Ridge, Tennessee 37831, USA}

\author{J.~Ross~Stewart}
\affiliation{ISIS Neutron and Muon Source, Rutherford Appleton
Laboratory, Didcot OX11 0QX, UK}

\author{Joseph~W.~Kolis}
\affiliation{Department of Chemistry, Clemson University, Clemson, South Carolina 29634-0973, USA}

\author{Adam~A.~Aczel}
\email{aczelaa@ornl.gov}
\affiliation{
	Neutron Scattering Division, Oak Ridge National Laboratory, Oak Ridge, Tennessee 37831, USA}

\begin{abstract}
We report single-crystal neutron spectroscopy and bulk characterization on hydrothermally grown Nd$_2$Sn$_2$O$_7$, revealing a dynamical moment fragmentation embedded within the all-in–all-out ordered state. The spectra show a nearly flat band with pinch-point-like momentum dependence, accompanied by dispersive branches that generate half-moon features across multiple Brillouin zones. These defining signatures are captured quantitatively by a minimal dipolar–octupolar spin Hamiltonian, demonstrating excellent agreement between experiment and theory. The higher flat-mode energy helps account for the absence of dynamical interference in prior muon spin relaxation $\mu$SR studies, while the lack of any photon-like excitation imposes strict constraints on the proposed Coulombic antiferromagnet scenario. Our results extend dynamical moment fragmentation to Nd$_2$Sn$_2$O$_7$ and identify it as a clean, tractable platform for quantitative exploration of emergent gauge-field physics and multipolar spin-wave dynamics in frustrated magnets.
\end{abstract}

\maketitle

\begin{figure*}[t] 
\centering
\includegraphics{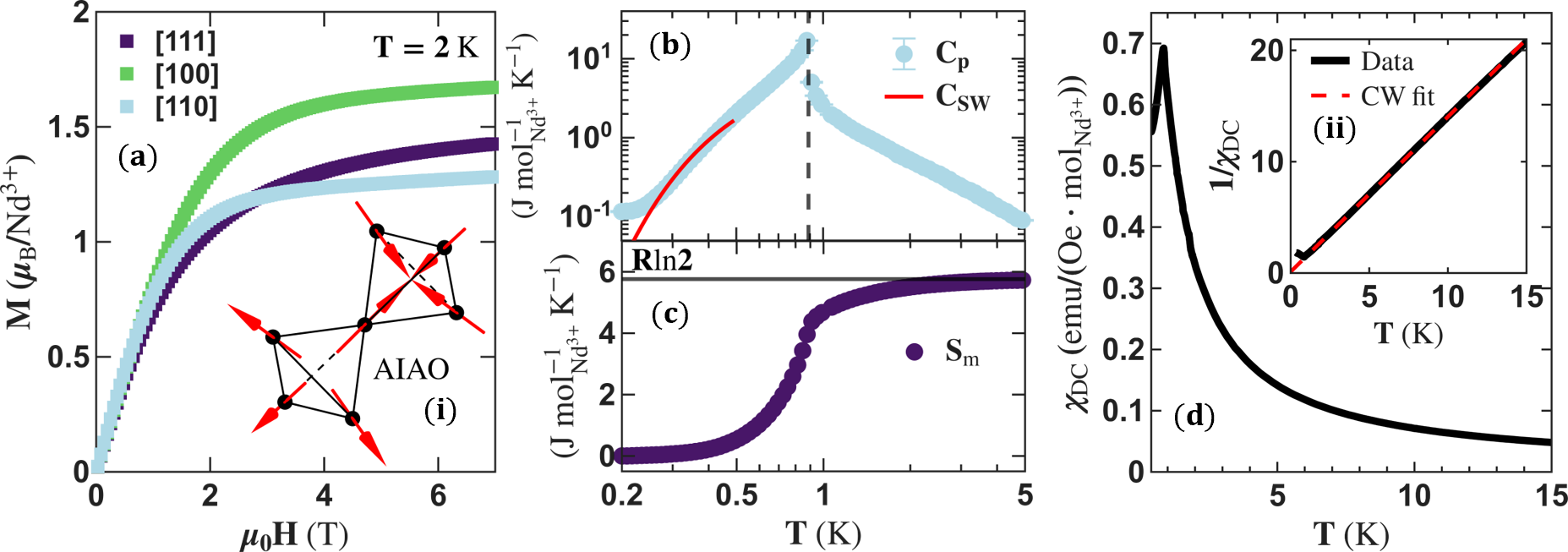}
\caption{(a) Magnetization $M(H)$ at $T = 2$~K for \nso\ measured along $[100]$, $[111]$, and $[110]$. Inset (i): schematic of the all-in–all-out (AIAO) order below $T_{\mathrm N}$. 
(b) Heat capacity $C_p(T)$ after subtracting the La$_2$Sn$_2$O$_7$ lattice contribution; the vertical dashed line marks $T_{\mathrm N} = 0.89(3)$~K; the red curve marks the calculated spin-wave contribution $C_{\mathrm{SW}}(T)$. (c) Magnetic entropy $S_{\mathrm{mag}}(T) = \int_0^T (C_p/T')\,\mathrm{d}T'$ recovering $R\ln2$ by $\sim$5~K. 
(d) DC susceptibility $\chi(T)$ of pulverized single crystals showing a peak at $T_{\mathrm N}$. Inset (ii): Curie–Weiss fit to $\chi^{-1}(T)$ over 2–10~K, yielding $\theta_{\mathrm{CW}} = -0.08(1)$~K and $\mu_{\mathrm{eff}} = 2.397(2)\,\mu_{\mathrm B}$/Nd$^{3+}$.}
\label{fig:Fig1}
\end{figure*}

\textit{Introduction--}Geometrical frustration on the pyrochlore lattice of corner-sharing tetrahedra gives rise to spin-ice (Coulombic) physics~\cite{Gardner2010RMP,Smith2025ARCMP,Balents2010_SpinLiquids,Savary2017_QSLreview}. In classical spin ice, strong local $\langle 111\rangle$ Ising anisotropy together with effective ferromagnetic nearest-neighbor exchange enforces the “two-in, two-out’’ rule on each tetrahedron~\cite{Bramwell2001,DenHertog2000}, producing an extensively degenerate manifold with Pauling entropy~\cite{ramirez1999zero} and dipolar correlations that yield pinch-point patterns in diffuse scattering~\cite{Fennell2009}. These results motivate the quantum spin-ice (QSI) scenario~\cite{Savary2012}, in which anisotropic exchanges induce tunneling between ice configurations and stabilize a $U(1)$ quantum spin liquid (QSL) with an emergent gauge field and gapless photon-like excitations.

Amid the intense search for materials that realize QSI physics, the dipolar–octupolar (DO) rare-earth pyrochlores R$_2$M$_2$O$_7$ (R = Ce, Nd, Sm; M = Zr, Hf, Sn, Ti, Pb)~\cite{Gaudet2019_Ce2Zr2O7_QSI_PRL,bhardwaj2022sleuthing,Gao2019Ce2Zr2O7,Smith2022U1pi,Smith2025_Ce2Zr2O7_PRX,gao2025neutron,Sibille2019_Ce2Sn2O7_octupole_NatPhys,Poree2025FractionalCe2Sn2O7,Yahne2024_Ce2Sn2O7_PRX,Poree2023Ce2Hf2O7,smith2025two,Xu2015_Nd2Zr2O7_CrystalField,Lhotel2015_DipoleOctupole_Nd2Zr2O7,Petit2016NZO,xu2016spin,benton2016quantum,Lhotel2018_DynamicKagomeIce,Xu2018_QSChains_Nd2Zr2O7,Xu2019_AnisotropicExchange_Nd2Zr2O7,Xu2020_OrderOut_Higgs_Nd2Zr2O7,Leger2021_SpinDynamics_Nd2Zr2O7,Leger2024_Disorder_NdPyro,anand2015observation,Anand2017_Nd2Hf2O7_muSR_INS,samartzis2022pinch,Bertin_2015_Nd2Sn2O7,DalmasDeReotier2017SlowSpinTunneling,hallas2015magnetic, swarnakar2017chemical, malkin2010static, singh2008manifestation, mauws2018dipolar, peccanha2019intermultiplet} have emerged as a particularly compelling family. Because the ground-state doublet of DO pyrochlores is typically well separated from the first crystal-electric-field (CEF) excitation by $\Delta_{\mathrm{CEF}}\gtrsim 100\,\mathrm{K}$, while the nearest-neighbor interactions are only $J\sim 1\,\mathrm{K}$, the low-energy physics is well captured by an effective pseudospin-$\tfrac{1}{2}$ description in terms of operators $\tau_i^{x,y,z}$~\cite{huang2014quantum}. In the local frame with $\hat{\mathbf z}\!\parallel\!\langle 111\rangle$ and $\hat{\mathbf y}$ along a twofold rotation axis, only $\tau_i^{z}$ has a nonzero magnetic-dipole matrix element, whereas $\tau_i^{x}$ and $\tau_i^{y}$ have leading octupolar character. From a symmetry standpoint, $\tau_i^{x}$ and $\tau_i^{z}$ transform as components of a dipolar vector under time reversal and $D_{3d}$ symmetry, while $\tau_i^{y}$ transforms as a component of the magnetic octupole tensor~\cite{huang2014quantum,Rau_Gingras_2019_FrustratedQuantumRareEarthPyrochlores}. Within the DO manifold, the symmetry-allowed nearest-neighbor exchange reduces, after an appropriate pseudospin rotation, to the compact “XYZ” Hamiltonian~\cite{huang2014quantum,benton2016quantum}:
\begin{equation}
\mathcal{H}^{\mathrm{DO}}_{\mathrm{XYZ}}
=\sum_{\langle ij\rangle}\!\left(
\tilde{J}_x\,\tilde{\tau}_i^{\tilde{x}}\tilde{\tau}_j^{\tilde{x}}
+\tilde{J}_y\,\tilde{\tau}_i^{\tilde{y}}\tilde{\tau}_j^{\tilde{y}}
+\tilde{J}_z\,\tilde{\tau}_i^{\tilde{z}}\tilde{\tau}_j^{\tilde{z}}
\right).
\label{HXYZ}
\vspace{-1em}
\end{equation}
Here, the pseudospin-\(\tfrac{1}{2}\) operators \(\tilde{\tau}_i^{\tilde{x},\tilde{y},\tilde{z}}\) are defined in a rotated local frame \((\tilde{x},\tilde{y},\tilde{z})\) obtained by a rotation by a material-dependent angle \(\vartheta\) about the local \(y\) axis; consequently \(\tilde{\tau}_i^{\tilde{y}}=\tau_i^{y}\) retains its octupolar character, whereas \(\tilde{\tau}_i^{\tilde{x}}\) and \(\tilde{\tau}_i^{\tilde{z}}\) are linear combinations of \(\tau_i^{x}\) and \(\tau_i^{z}\) that transform as magnetic dipole components. Theoretical studies~\cite{Li2017_SEU1_DOpyro,Yao2020_PyroU1_MixedSym,Benton2020_GS_Phase_DO,Patri2020_Magnetostriction_QSI} show that DO pyrochlores can host a variety of symmetry-enriched \(U(1)\) QSL phases; possible realizations have recently been reported in Ce-based pyrochlores~\cite{Gaudet2019_Ce2Zr2O7_QSI_PRL,bhardwaj2022sleuthing,Gao2019Ce2Zr2O7,Smith2022U1pi,Smith2025_Ce2Zr2O7_PRX,gao2025neutron,Sibille2019_Ce2Sn2O7_octupole_NatPhys,Poree2025FractionalCe2Sn2O7,Yahne2024_Ce2Sn2O7_PRX,Poree2023Ce2Hf2O7,smith2025two}.

By contrast, Nd-based pyrochlores occupy a different region of the DO phase diagram and stabilize the all-in–all-out (AIAO) antiferromagnetic state, in which the four moments on each tetrahedron point either all toward or all away from the center, alternating on neighboring tetrahedra [Fig.~\ref{fig:Fig1}(i)]. Yet this order is far from trivial: geometric frustration together with the peculiar symmetry of the DO doublet enables \emph{moment fragmentation}~\cite{BrooksBartlett2014_MMF,Lhotel2020_FragmentationReview,benton2016quantum}, wherein the dipolar magnetization field decomposes into divergence-full and divergence-free components. In the original proposal of Ref.~\cite{BrooksBartlett2014_MMF}, which we refer to here as \emph{static moment fragmentation}, both components reside in the ground state and correspond to the coexistence of long-range order with a Coulomb-phase spin liquid. By contrast, in Nd-based dipole–octupole pyrochlores~\cite{benton2016quantum} the mechanism arises from multipolar couplings in the effective XYZ model [Eq.~\ref{HXYZ}], where the divergence-full component forms the static AIAO order while the divergence-free sector appears within finite-energy excitations. We refer to this scenario as \emph{dynamical moment fragmentation}. A prototypical realization is \nzo~\cite{Xu2015_Nd2Zr2O7_CrystalField,Lhotel2015_DipoleOctupole_Nd2Zr2O7,Petit2016NZO,xu2016spin,benton2016quantum,Lhotel2018_DynamicKagomeIce,Xu2018_QSChains_Nd2Zr2O7,Xu2019_AnisotropicExchange_Nd2Zr2O7,Xu2020_OrderOut_Higgs_Nd2Zr2O7,Leger2021_SpinDynamics_Nd2Zr2O7,Leger2024_Disorder_NdPyro}, in which neutron diffraction and spectroscopy reveal antiferromagnetic Bragg peaks coexisting with a gapped flat band at $\sim\!70~\upmu$eV whose structure factor displays the characteristic pinch-point pattern~\cite{Petit2016NZO}. Theory~\cite{benton2016quantum,Xu2019_AnisotropicExchange_Nd2Zr2O7} traces these phenomena to the XYZ model: a ferromagnetic $\tilde{J}_{\tilde{z}}<0$ in the local frame selects AIAO order of the $\tilde{\tau}^{\tilde{z}}$ component, while an antiferromagnetic $\tilde{J}_{\tilde{x}}>0$ frustrates the transverse $\tilde{\tau}^{\tilde{x}}$ fluctuations. For $\tilde{J}_{\tilde{x}}<3|\tilde{J}_{\tilde{z}}|$, the ground state hosts long-range AIAO order, and the dipolar moment naturally partitions into static and fluctuating components,
\begin{equation}
\mathbf{m}_i
= g_z\,\uB\!\left[\cos\vartheta\,\tilde{\tau}_i^{\tilde{z}}
+ \sin\vartheta\,\tilde{\tau}_i^{\tilde{x}}\right]\hat{z}_i,
\label{moment_frag}
\end{equation}
where $g_z$ is the longitudinal $g$ factor and $\hat{z}_i$ the local easy axis at site $i$. The $\tilde{\tau}_i^{\tilde{z}}$ component forms the static ordered moment $m_{\mathrm{ord}}=1.26(2)~\uB/\mathrm{Nd}^{3+}$~\cite{Xu2015_Nd2Zr2O7_CrystalField}, whereas $\tilde{\tau}_i^{\tilde{x}}$ remains dynamic and gives rise to the flat band with spin-ice–like correlations. Upon warming just above the N\'eel temperature $T_{\mathrm{N}}\!\approx\!0.4$~K, this band softens to the elastic line, realizing a gapless Coulomb phase and signaling the proximity of \nzo\ to the $U(1)$ QSL regime~\cite{Xu2020_OrderOut_Higgs_Nd2Zr2O7,Leger2021_SpinDynamics_Nd2Zr2O7}.

Establishment of dynamical moment fragmentation in \nzo\ prompts the question of its prevalence across the Nd pyrochlore family. Neutron spectroscopy on \nho~\cite{anand2015observation,Anand2017_Nd2Hf2O7_muSR_INS,samartzis2022pinch} and Nd$_2$ScNbO$_7$~\cite{mauws2021magnetic,scheie2021beyond} confirms AIAO order together with dynamical spin-ice-like correlations, the hallmarks of moment fragmentation. In contrast, the divergence free component, and thus dynamical fragmentation, has been argued to be absent in \nso~\cite{Bertin_2015_Nd2Sn2O7}, which enters an AIAO phase below $T_{\mathrm{N}}\approx 0.91~\mathrm{K}$ with a comparatively large ordered moment $m_{\mathrm{ord}}=1.708(3)\,\uB/\mathrm{Nd}^{3+}$ (see Table~\ref{Tab:Compare}). The central experimental claim is the observation of spontaneous muon spin precession in zero field muon spin relaxation ($\mu$SR) below $T_{\mathrm{N}}$, absent in analogous measurements on \nzo~\cite{xu2016spin} and \nho~\cite{Anand2017_Nd2Hf2O7_muSR_INS}; this has been taken to indicate that no dynamic internal fields exist on the $\mu$SR timescale to obscure the static order in \nso. Taken together with the observation that $C_p\propto T^{3}$ below $T_{\mathrm{N}}$, these results motivated the proposal that \nso\ realizes a Coulombic antiferromagnet (CAF), stabilized by further neighbor interactions and characterized by long range spin ice entanglement~\cite{Chen2023_CAF}. The CAF phase is predicted to exhibit distinctive spectroscopic signatures, including gapped monopole continuums and gapless, linearly dispersing photon-like excitations. Following Ref.~\cite{Bertin_2015_Nd2Sn2O7}, work on Nd$_2$GaSbO$_7$~\cite{gomez2021absence} likewise finds no conclusive evidence of dynamical moment fragmentation within the AIAO state; instead a gapped magnetic mode appears at $\hbar\omega\approx 0.25~\mathrm{meV}$ without the $Q$-dependence characteristic of spin-ice correlations.

Despite suggestive bulk characterizations and $\mu$SR results on \nso~\cite{Bertin_2015_Nd2Sn2O7,Bertin_2015_thesis}, these probes cannot distinguish a fragmented AIAO state from a conventional AIAO phase, nor can they establish the proposed CAF scenario. A decisive test requires single-crystal neutron scattering, which directly probes the dynamical structure factor and can reveal either the pinch-point flat band characteristic of moment fragmentation or the photon/monopole excitations expected for a CAF. Such experiments were long impeded by the difficulty of growing sizable single crystals of \nso, a challenge recently overcome by advances in hydrothermal growth methods~\cite{powell2019hydrothermal}. Here we report a bulk characterization and single-crystal neutron spectroscopy study of \nso. These measurements reveal the spectroscopic signatures of moment fragmentation and delineate the low-temperature magnetic dynamics of \nso, thereby distinguishing this scenario from the Coulombic antiferromagnet proposal and clarifying the extent of dynamical moment fragmentation across the Nd pyrochlore family.

\begin{figure*}[t] 
\centering
\includegraphics{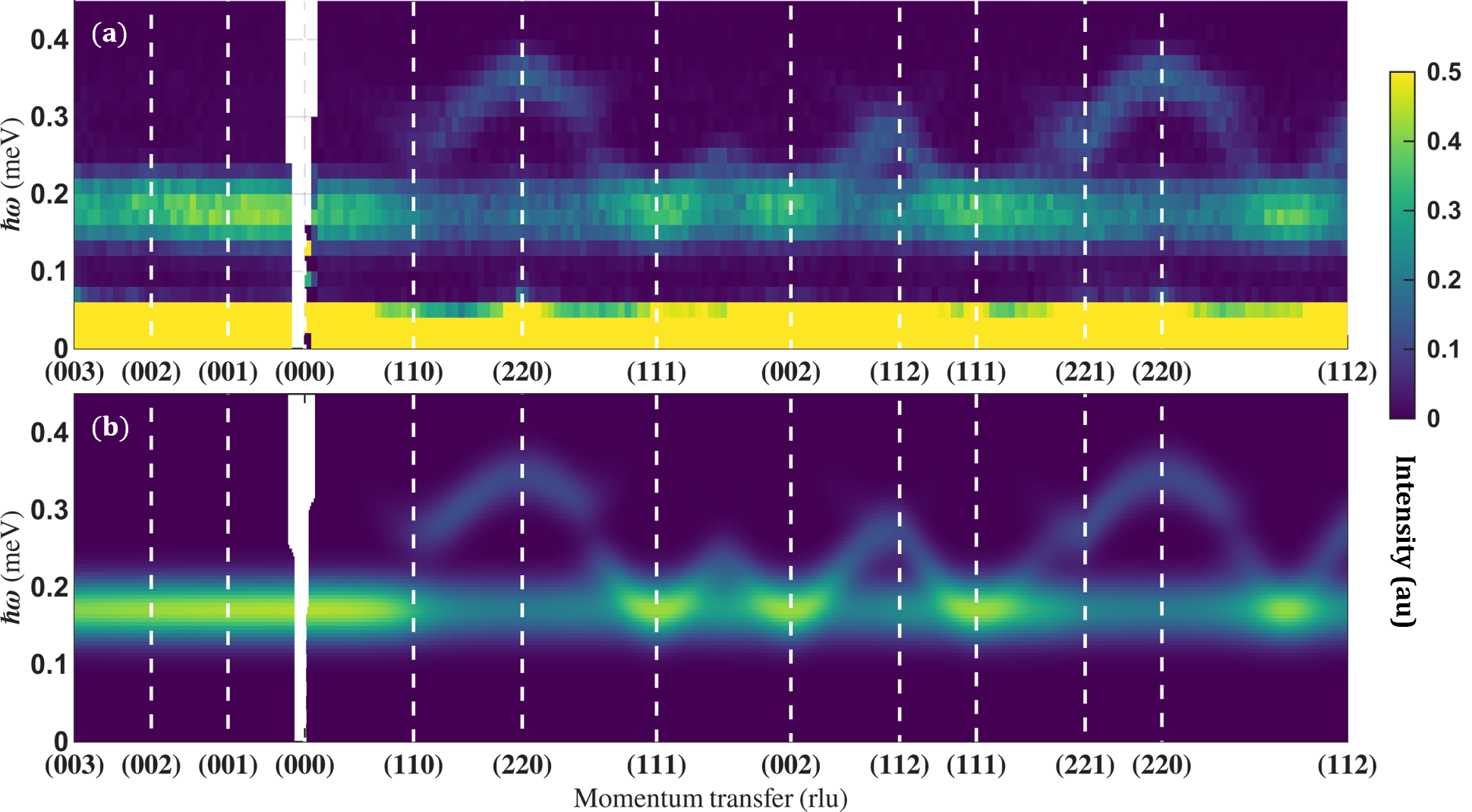}
\caption{(a) False color map of inelastic scattering intensity $S(\mathbf{Q},\omega)$ versus energy transfer $\hbar\omega$ and momentum transfer $Q$ along a path connecting the labeled high symmetry points. Data were taken at $T=0.25$~K with incident energy $E_i=2.19$~meV on the LET spectrometer (elastic resolution full width at half maximum (FWHM) $\approx0.07$ meV), integrated over a perpendicular momentum window of $\pm 0.1$~\AA$^{-1}$, and not symmetrized. (b) Simulated intensity from the spin-wave model of Ref.~\cite{benton2016quantum} with parameters $(\tilde{J}_x,\tilde{J}_y,\tilde{J}_z)=(0.10(2),\,0.00(1),\,-0.075(2))~\mathrm{meV}$, convolved with the LET energy resolution and scaled by a single global factor to match the data.}
\label{fig:Fig2}
\end{figure*}

\begin{figure}[t]
\centering
\includegraphics{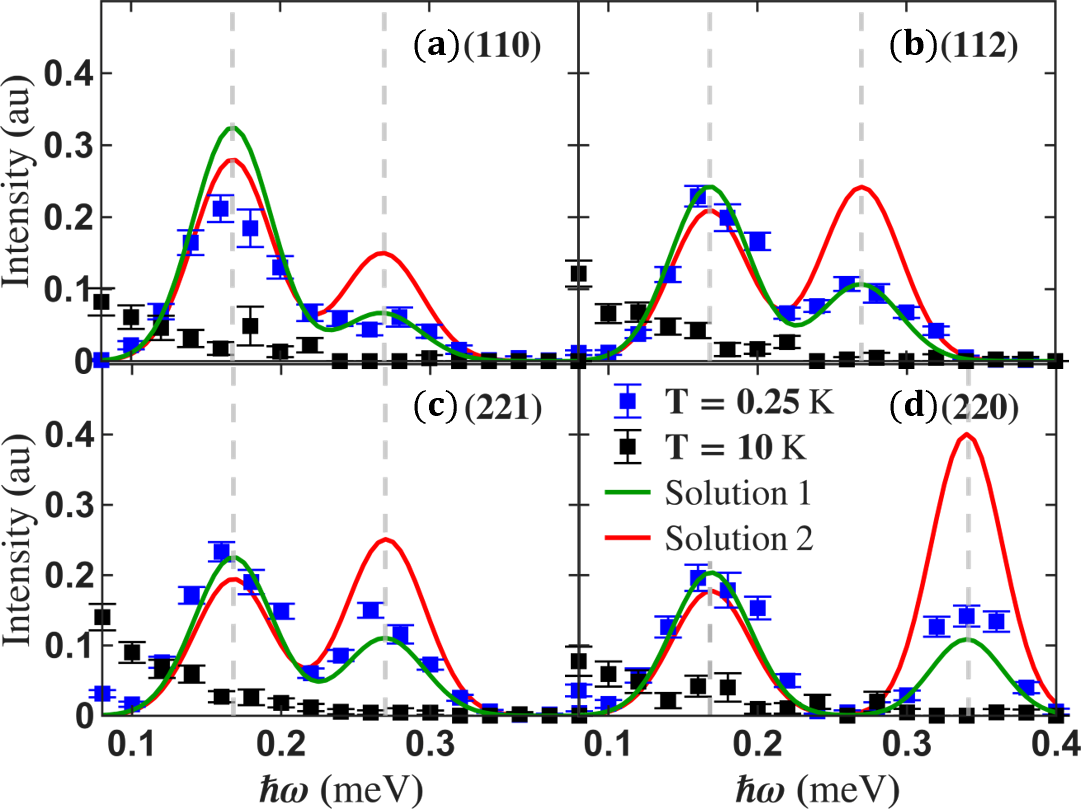}
\caption{(a–d) Constant-$Q$ energy cuts at the indicated points, integrated over $\pm0.1$~reciprocal lattice unit (rlu) along each principal direction and measured with $E_i=2.19$~meV. Blue/black: $T=0.25$~K/$10$~K. Green/red: simulations for Solution~1/Solution~2, averaged over the same $Q$-windows as the data, convolved with the Low Energy Transfer (LET) energy resolution, and scaled by a single global factor (per solution) to match the total intensity across panels. Experimental spectra are averaged over symmetry-equivalent points in the $(HHL)$ plane using space group $Fd\bar{3}m$. Gray dashed lines mark the calculated spin-wave energies.}
\label{fig:Fig3}
\end{figure}

\begin{figure*}[t] 
\centering
\includegraphics{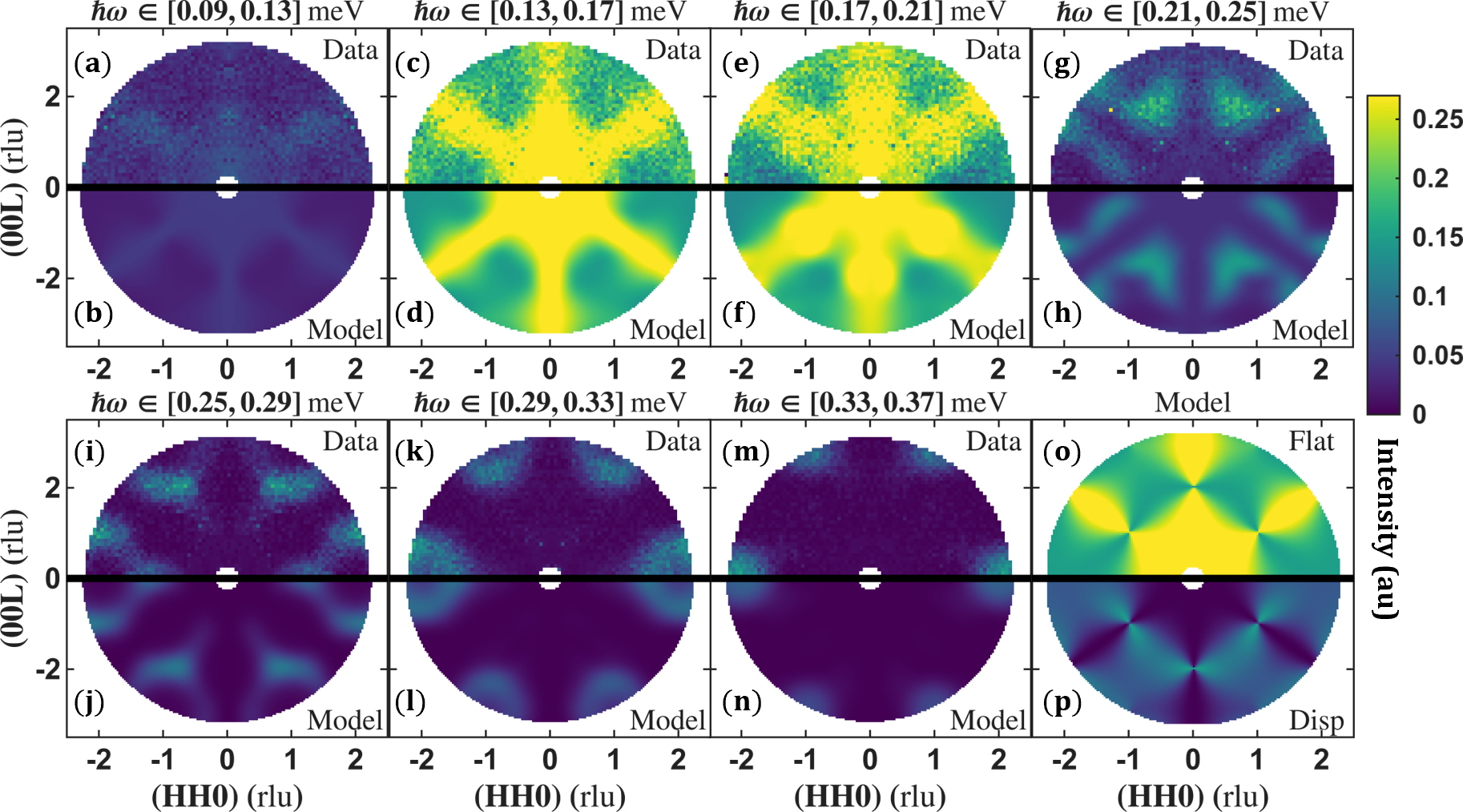}
\caption{(a,c,e,g,i,k,m) Constant energy $Q$ maps collected at $T=0.25$~K with $E_i=2.19$~meV in the $(HHL)$ plane. Data are integrated along $[K\bar{K}0]$ with $K\in[-0.07,0.07]$~rlu, and symmetrized in the $(HHL)$ plane using space group $Fd\bar{3}m$. (b,d,f,h,j,l,n) Simulated $Q$ maps from the spin wave model of Ref.~\cite{benton2016quantum} with parameters $(\tilde{J}_x,\tilde{J}_y,\tilde{J}_z)=(0.10(2)),0.00(1),-0.075(2)))~\mathrm{meV}$, convolved with the LET resolution and scaled by a single global factor to match the data. (o,p) $Q$-dependence of the simulated intensity for the flat band (o) and the dispersive branches (p).}
\label{fig:Fig4}
\end{figure*}

\textit{Results--} Single crystal \nso\ samples were prepared \textit{via} a hydrothermal method. Magnetization $M(H)$ was measured along $[100]$, $[111]$, and $[110]$ directions [Fig.~\ref{fig:Fig1}(a)]. The saturation moments are ordered as $m_{100}^{\mathrm{sat}} > m_{111}^{\mathrm{sat}} > m_{110}^{\mathrm{sat}}$, consistent with expectations for a magnetic pyrochlore with easy axis anisotropy~\cite{Fukazawa2002}. Heat capacity after subtracting the lattice contribution [Fig.~\ref{fig:Fig1}(b)], shows a $\lambda$ type peak at $T_{\mathrm{N}}=0.89(3)$~K. The integrated magnetic entropy $S_{\mathrm{mag}}(T)$ approaches the full $R\ln 2$ of the ground state doublet by $\sim 5$~K [Fig.~\ref{fig:Fig1}(c)]. The dc susceptibility of pulverized single crystals of \nso\ exhibits a clear anomaly at $T_{\mathrm{N}}$ [Fig.~\ref{fig:Fig1}(d)]; a Curie Weiss fit over 2--10 K yields $\theta_{\mathrm{CW}}=-0.08(1)$~K and an effective moment $\mu_{\mathrm{eff}}=2.397(2)\,\uB/\mathrm{Nd^{3+}}$ [Fig.~\ref{fig:Fig1}(ii)]. For inelastic neutron spectroscopy, we measured a co-aligned, coarse-grained assembly on the Low Energy Transfer (LET) cold-neutron chopper spectrometer at the ISIS Neutron and Muon Source (Rutherford Appleton Laboratory, UK). Details of the experimental methods are provided in the Supplemental Material.

An overview of the inelastic magnetic response of \nso\ at $T=0.25$~K, deep in the AIAO phase below $T_{\mathrm{N}}$, is shown in Fig.~\ref{fig:Fig2}(a). Magnetic Bragg peaks at $(220)$ and $(113)$ appear on the elastic line (see Fig.~S2 of the Supplemental Material), consistent with the expected AIAO magnetic structure of the divergence-full $\tilde{\tau}^{\tilde{z}}$ component~\cite{benton2016quantum} and in agreement with previous powder measurements~\cite{Bertin_2015_Nd2Sn2O7}. The inelastic channel displays gapped spin-wave excitations comprising a nearly flat band and dispersive branches, whose momentum and energy dependences closely resemble those reported for \nzo~\cite{Petit2016NZO,Xu2019_AnisotropicExchange_Nd2Zr2O7} and \nho~\cite{samartzis2022pinch}. Within the dynamical fragmentation framework~\cite{benton2016quantum}, the nearly flat band corresponds to the divergence-free sector, while the dispersive branches arise from the divergence-full sector of the fluctuating $\tilde{\tau}^{\tilde{x}}$ component. No monopole continuum or photon-like mode is observed within the accessible energy range $\hbar\omega \gtrsim 0.09$~meV.

To model the spectra we employ the spin-wave framework of Ref.~\cite{benton2016quantum}. The exchange set $(\tilde{J}_x,\tilde{J}_y,\tilde{J}_z)$ is obtained by fitting the calculated energies to the measured gaps of the flat band, $\Delta_1=0.168(2)$~meV, the dispersive branches at the zone boundary $(100)$, $\Delta_2=0.270(4)$~meV, and the highest energy at the zone center $(000)$, $\Delta_3=0.341(5)$~meV. This procedure yields two symmetry-related solutions: Solution~1, $(0.10(2),\,0.00(1),\,-0.075(2))$~meV, and Solution~2, obtained by exchanging $\tilde{J}_x$ with $\tilde{J}_y$. Constant-$Q$ energy cuts at representative high-symmetry points [Fig.~\ref{fig:Fig3}], benchmarked against resolution-convolved simulations, show that only Solution~1 reproduces the relative spectral weights of the flat and dispersive modes. The simulated lineshapes match the experimental, resolution-limited linewidths, implying no detectable magnon damping. Simulated inelastic spectra based on Solution~1 are shown in Fig.~\ref{fig:Fig2}(b), and constant-energy $Q$ maps in Fig.~\ref{fig:Fig4}. We find excellent agreement between the data and our simulations for both the dispersions and the structure factors across multiple Brillouin zones. The flat band at $\Delta_{1}=0.168(2)\,$meV exhibits the characteristic pinch-point-like $Q$ dependence~\cite{Petit2016NZO,benton2016quantum} [Figs.~\ref{fig:Fig4}(a)–(d)], while the dispersive branches immediately above it display the characteristic ‘half-moon’ features~\cite{Yan2018_HalfMoons_PRB,samartzis2022pinch} [Figs.~\ref{fig:Fig4}(e)–(h)]. The dispersive mode with higher energy reaches its band maximum at $\Delta_{3}=0.341(5)\,$meV at the zone centers $(220)$ and $(113)$ [Figs.~\ref{fig:Fig4}(i)–(n)].

With the exchange parameters determined, the rotation angle $\vartheta$, which sets the relative degrees of the static and fluctuating moment components in Eq.\ref{moment_frag}, can be estimated from $m_{\mathrm{ord}}/m_{\mathrm{sat}}
=\cos\vartheta\langle\tilde{\tau}^{\tilde{z}}\rangle/S$, with $S=\tfrac12$~\cite{benton2016quantum,Xu2019_AnisotropicExchange_Nd2Zr2O7}. Using the reported ordered moment $m_{\mathrm{ord}}=1.708(3)\,\uB$~\cite{Bertin_2015_Nd2Sn2O7} and taking $m_{\mathrm{sat}}\approx\mu_{\mathrm{eff}}=2.397(2)\,\uB$ for an Ising-like ground state doublet~\cite{gomez2021absence}, we obtain $\vartheta=0.74(3)$~rad for \nso. The Curie Weiss temperature calculated from $\vartheta$~\cite{benton2016quantum} is $\theta_{\mathrm{CW}}^{\mathrm{cal}}=0.03(7)$~K, in reasonable agreement with our experimental fit. The calculated spin-wave contribution to the heat capacity, $C_{\mathrm{SW}}$, is shown in Fig.~\ref{fig:Fig1}(b). The slight upturn of $C_p$ below $\sim$0.25~K may stem from a nuclear contribution, similar to the behavior reported for \nzo~\cite{Xu2019_AnisotropicExchange_Nd2Zr2O7}. Details of the spin-wave model and parameters determination are provided in the Supplemental Material.

\begin{table}[b]
\label{tab:fivebyfive}
\begin{ruledtabular}
\begin{tabular}{lcccc}
\textrm{Sample} & $m_{\mathrm{ord}}~(\uB)$ & $m_{\mathrm{sat}}~(\uB)$ & $\Delta_1~\mathrm{(meV)}$ & $\vartheta~\mathrm{(rad)}$ \\
\hline\hline
\nzo & 1.26(2)\cite{Xu2015_Nd2Zr2O7_CrystalField} & 2.50$^*$\cite{Xu2019_AnisotropicExchange_Nd2Zr2O7} & 0.075(4)\cite{Xu2019_AnisotropicExchange_Nd2Zr2O7} & 0.98(3)\cite{Xu2019_AnisotropicExchange_Nd2Zr2O7} \\
\nho & 0.62(1)\cite{anand2015observation} & 2.50\cite{anand2015observation} & 0.094\cite{samartzis2022pinch} & 1.265\cite{samartzis2022pinch} \\
Nd$_2$ScNbO$_7$ & 2.2(4)\cite{mauws2021magnetic} & 2.25(25)\cite{mauws2021magnetic} & 0.065(15)\cite{mauws2021magnetic} & - \\
Nd$_2$GaSbO$_7$  & 1.59(5)\cite{gomez2021absence}  & 2.37(1)\cite{gomez2021absence} & 0.253(6)\cite{gomez2021absence} & - \\
\nso & 1.708(3)\cite{Bertin_2015_Nd2Sn2O7} & 2.397(2)$^*$ & 0.168(2) & 0.74(3) \\
\end{tabular}
\end{ruledtabular}
\caption{Comparison of ordered moment $m_{\mathrm{ord}}$, saturation moment $m_{\mathrm{sat}}$, flat-mode gap $\Delta_1$, and rotation angle $\vartheta$ for selected Nd pyrochlores. Entries for $m_{\mathrm{sat}}$ marked with an asterisk ($^{*}$) are inferred from magnetic susceptibility rather than a CEF analysis.}
\label{Tab:Compare}
\end{table}

\textit{Discussion--} The bulk thermodynamic measurements on our hydrothermally grown single-crystal \nso, 
in particular the heat capacity $C_p$ below $T_{\mathrm{N}}$ 
[Fig.~\ref{fig:Fig1}(b)], are consistent with the powder results reported in 
Ref.~\cite{Bertin_2015_Nd2Sn2O7}.  
However, the calculated spin-wave heat capacity $C_{\mathrm{SW}}$ accounts for most of the 
measured $C_p$ in the range $0.2$--$0.5$~K, a regime previously described by 
$C_p \propto T^{3}$~\cite{Bertin_2015_Nd2Sn2O7} and attributed to a linearly dispersing 
photon mode~\cite{Bertin_2015_Nd2Sn2O7,Chen2023_CAF}.  

Consistent with this picture, our neutron spectra establish dynamical moment fragmentation in \nso\ 
below $T_{\mathrm N}$. Elastic intensity at $(220)$ and $(113)$ on the elastic line is consistent with the AIAO order of the $\tilde{\tau}^{\tilde{z}}$ component, while the inelastic channel reveals gapped spin-wave excitations arising from the $\tilde{\tau}^{\tilde{x}}$ sector. The flat band at $\Delta_{1}=0.168(2)$~meV lies well above the dynamical window of zero-field $\mu$SR (estimated to be $\lesssim 0.05$~meV~\cite{hatt2025cluster,hillier2022muon}), which likely accounts for the absence of $\mu$SR-detected dynamical interference with 
static order in \nso, in contrast to \nzo~\cite{xu2016spin} and 
\nho~\cite{Anand2017_Nd2Hf2O7_muSR_INS} where the substantially smaller $\Delta_{1}$ renders such effects observable.  
Taken together, our neutron spectra rule out CAF excitations within the accessible energy range $\hbar\omega \gtrsim 0.09$~meV, while the prior $\mu$SR measurements likewise 
exclude low-energy modes with $\hbar\omega\lesssim 0.05$~meV. These combined constraints strongly limit the energy window available for CAF excitations and effectively rule out a dominant CAF phase in \nso.

Linear spin-wave calculations within the dynamical fragmentation framework~\cite{benton2016quantum} allow a quantitative determination of the exchange parameters $(\tilde{J}_x,\tilde{J}_y,\tilde{J}_z) = (0.10(2),\,0.00(1),\,-0.075(2))~\mathrm{meV}$. The near-vanishing $\tilde{J}_y$ disfavors any octupolar order of the $\tau^y$ component, in contrast to proposals for \cso~\cite{Sibille2019_Ce2Sn2O7_octupole_NatPhys,Poree2025FractionalCe2Sn2O7} and \cho~\cite{Poree2023Ce2Hf2O7}. Within this description, the observed excitations correspond to multipolar spin-wave modes of the dipole–octupole doublet, while dynamical moment fragmentation refers to the emergent decomposition of the dipolar magnetization field into divergence-full and divergence-free sectors manifested in their structure factors. The resulting simulations reproduce both the dispersions and the structure factors across multiple Brillouin zones with excellent fidelity. The calculated flat band displays the expected dynamical spin-ice correlations [Fig.\ref{fig:Fig4}(o)]~\cite{benton2016quantum}, while the broadened pinch-point pattern observed experimentally [Fig.\ref{fig:Fig4}(a,c)] arises from finite energy resolution and contributions from the nearby dispersive mode [Fig.\ref{fig:Fig4}(p)] close to the pinch-point positions. Higher-resolution probes (e.g. backscattering or neutron spin echo) will be essential to test whether a small CAF component coexists with the AIAO order and to search for a proximate ferromagnetic Coulomb phase just above $T_{\mathrm{N}}$, motivated by the case of \nzo~\cite{Xu2020_OrderOut_Higgs_Nd2Zr2O7,Leger2021_SpinDynamics_Nd2Zr2O7}.

For \nzo, pronounced sample dependence has been reported~\cite{Leger2021_SpinDynamics_Nd2Zr2O7,Leger2024_Disorder_NdPyro}. In our \nso\ crystal we observe a weak Bragg reflection at the nominally forbidden $(002)$ position at $T=0.25$~K and $10$~K, which is most likely associated with extrinsic effects such as multiple scattering~\cite{Baroudi2015_SymmetryLightStuffing} (see Figs.~S2 and S3 in the Supplemental Material). Within our experimental sensitivity we do not observe any discernible influence of structural imperfections on the magnetic excitations: the inelastic spectra are quantitatively reproduced by the disorder-free spin-wave model. We note that magnetic properties of pyrochlore compounds can in general be sensitive to stoichiometry and structural disorder; however, the consistency between the measured spectra and the minimal Hamiltonian suggests that such effects do not play a dominant role in the present sample.

Beyond the spin-ice context, our results place \nso\ within a broader class of multipolar quantum magnets. Recent work on FeI$_2$ has shown that strong spin--orbit coupling can mix dipolar and quadrupolar channels, yielding collective modes that lie beyond conventional spin-wave theory~\cite{bai2021hybridized}. Similarly, in UO$_2$, the coexistence of dipolar and quadrupolar order supports dispersive multipolar excitations that acquire neutron visibility through dipole–multipole mixing~\cite{carretta2010quadrupolar}. A similar mechanism operates in Nd-based dipole–octupole pyrochlores: in \nso, the 
octupolar component of the CEF doublet mixes with the dipolar sector, yielding AIAO order and gapped modes of mixed multipolar character. The resulting excitation spectrum cannot 
be understood within a purely dipolar picture, underscoring the broader significance of \nso\ as a platform for probing multipolar quasiparticles in spin–orbit–entangled magnets.

Overall, our findings establish \nso\ as a platform in which AIAO order supports dynamical moment fragmentation arising from well-characterized multipolar interactions. By establishing the relevant energy and exchange scales, this work expands the landscape of Nd pyrochlores where emergent gauge phenomena and exotic quasiparticles can be probed 
with quantitative precision.

\textit{Data availability—}
The neutron scattering data supporting this study are openly available from the STFC ISIS Neutron and Muon Source~\cite{Luo2024_Nd2Sn2O7_data}. 
Processed data corresponding to all figures in the manuscript are available online~\cite{Luo2024_Nd2Sn2O7_processed}.

\textit{Acknowledgements--} We thank Gang Chen and Arun Paramekanti for useful discussion. The synthesis, crystal growth, and crystallographic studies were performed at Clemson University and supported by DoE BES Award Number DE-SC0020071. B.R.O. (bulk property measurements and analysis, crystallographic studies) gratefully acknowledges support from the U.S. Department of Energy (DOE), Office of Science, Basic Energy Sciences, Materials Sciences and Engineering Division. Y.L., J.A.M.P., and A.A.A. acknowledge support from the U.S. Department of Energy (DOE), Office of Science, Basic Energy Sciences, Scientific User Facilities Division. We also thank the SNS X-ray Laboratory at ORNL for access to the Laue instrument used in sample co-alignment. Experiments at the ISIS Neutron and Muon Source were carried out under beamtime allocation RB2410211 from the Science and Technology Facilities Council (STFC). 

\clearpage

\begin{center}
\textbf{\large Supplemental Material for \\[3pt] ``Dynamical moment fragmentation in the all-in all-out pyrochlore \texorpdfstring{Nd$_2$Sn$_2$O$_7$}{Nd2Sn2O7}"}
\end{center}

\renewcommand{\thesection}{S\arabic{section}}
\renewcommand{\thesubsection}{S\arabic{section}.\arabic{subsection}}
\setcounter{figure}{0}
\renewcommand\thefigure{S\arabic{figure}}
\renewcommand{\theequation}{S\arabic{equation}}
\setcounter{equation}{0}
\renewcommand{\thetable}{S\arabic{table}}
\setcounter{table}{0}

\section{Methods and Sample Information}

\begin{figure*}[bt] 
\includegraphics{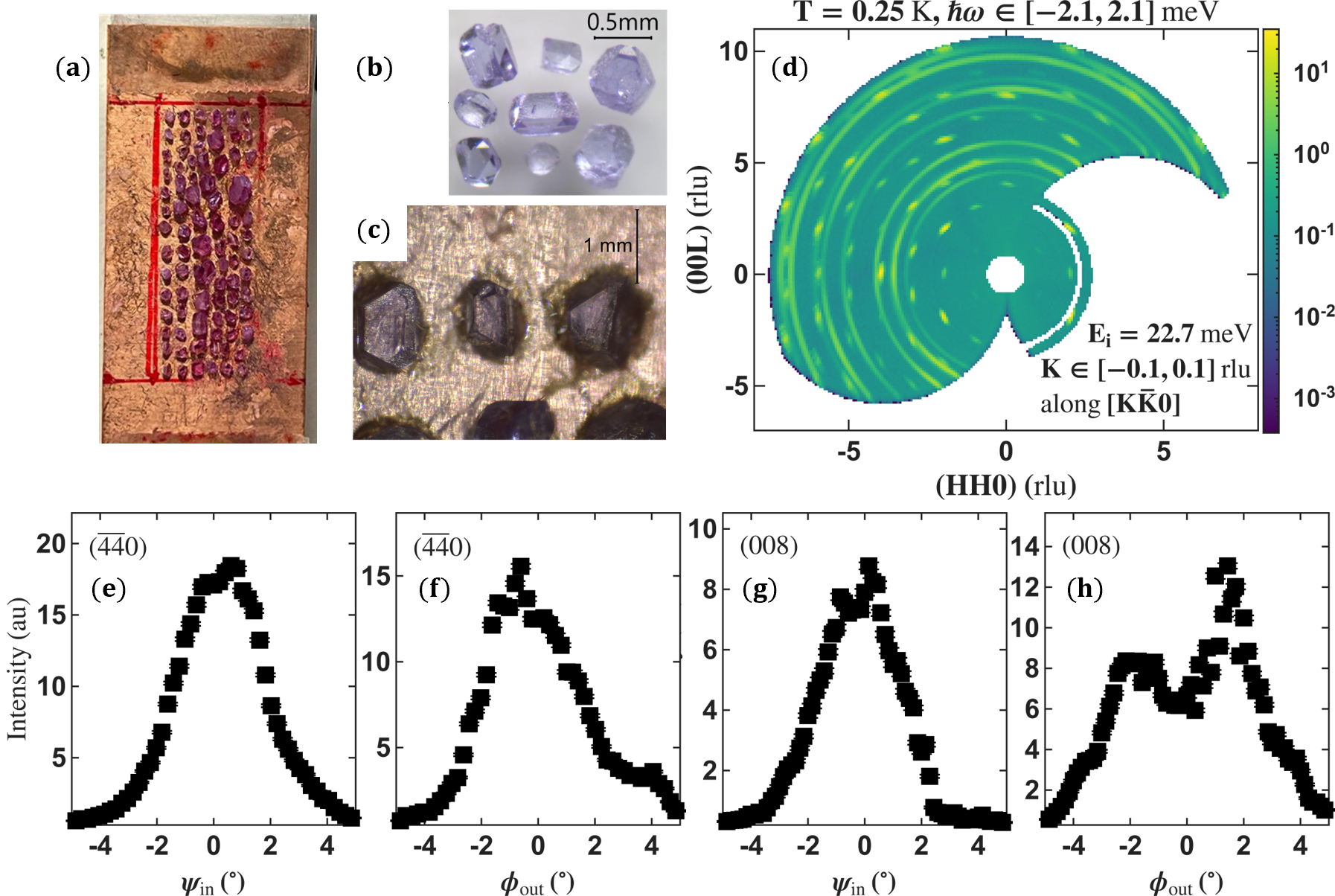}
\caption{(a) Photograph of a portion of the co-aligned single-crystal \nso\ array (total mass $\sim$0.93~g), with the vertical axis approximately aligned along the $[1\bar{1}0]$ crystallographic direction. (b) Optical microscopy image of a representative single crystal. (c) Optical microscopy image of crystals affixed to the copper plate within the co-aligned array shown in (a). (d) False-color intensity map of the $(HHL)$ plane of the co-aligned \nso\ crystal array, collected at $T=0.25$~K on the cold neutron multi-chopper spectrometer LET with incident energy $E_i = 22.7$~meV. The map displays the elastic channel integrated over energy transfer $[-2.1,2.1]$~meV and within $\pm0.1$~rlu along the $[K\bar{K}0]$ direction. The data reveal a single, slightly broadened crystal grain with an estimated mosaicity of $\sim$5$^\circ$ (FWHM), further illustrated in (e--h). (e--h) Transverse cuts through the Bragg peaks $\mathbf{Q}=(\bar{4}\bar{4}0)$ and $\mathbf{Q}=(008)$ along the in-plane directions $[001]$ and $[110]$, respectively, as well as the out-of-plane direction $[K\bar{K}0]$, measured at $T=0.25$~K. Horizontal axes are expressed in angular units $\psi_{\mathrm{in}}$ (in-plane) and $\phi_{\mathrm{out}}$ (out-of-plane), normalized to $|\mathbf{Q}|$. Data are integrated within $\pm0.09$~\AA$^{-1}$ in the perpendicular $\mathbf{Q}$ directions.}
\label{fig:SIFig1}
\end{figure*}

\begin{figure*}[tb] 
\includegraphics{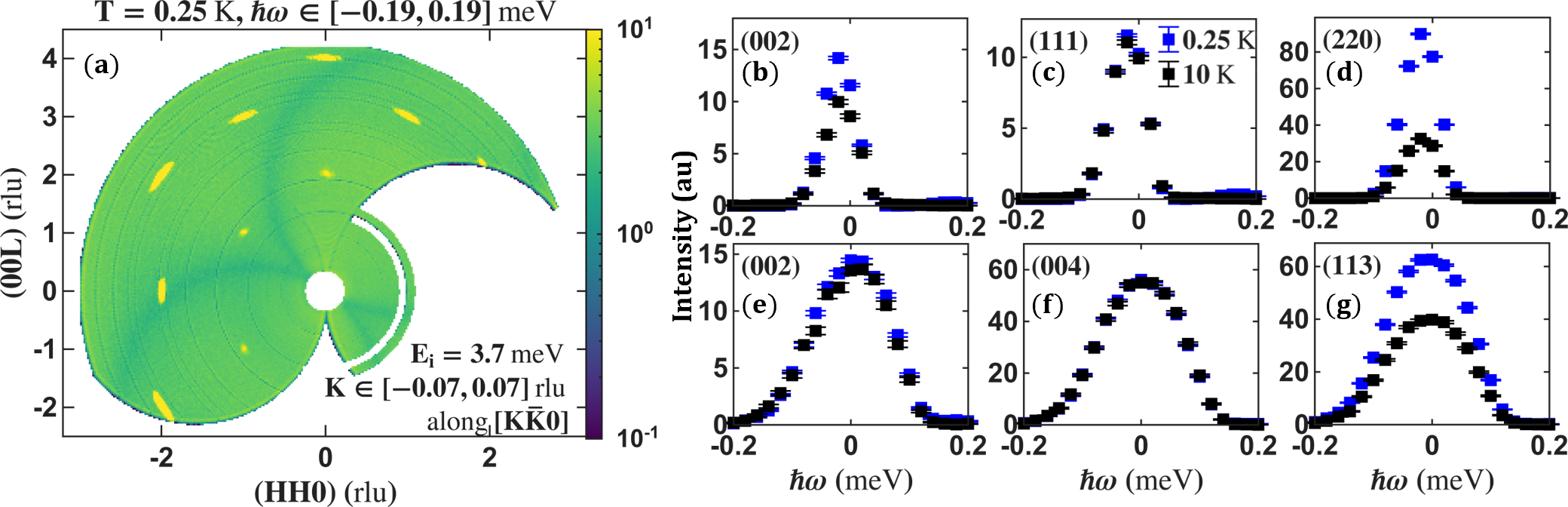}
\caption{(a) False-color intensity map of the $(HHL)$ plane of the co-aligned \nso\ crystal array, collected at $T=0.25$~K on the cold neutron multi-chopper spectrometer LET with incident energy $E_i=3.7$~meV. The map shows the elastic channel integrated over an energy range of $[-0.19,0.19]$~meV and within $\pm0.07$~rlu along the $[K\bar{K}0]$ direction. (b--g) Constant-$Q$ energy cuts at the positions indicated in each panel, integrated over $[-0.1,0.1]$~rlu along each principal reciprocal-space direction. Blue curves correspond to data collected at $T=0.25$~K and black curves to $T=10$~K. Panels (b--d) were measured with $E_i=2.19$~meV, while panels (e--g) were measured with $E_i=3.7$~meV. Panels (b--g) show data from Bragg peaks in the $(HHL)$ plane, symmetrized using the $Fd\bar{3}m$ space-group symmetry.
}
\label{fig:SIFig2}
\end{figure*}

\begin{figure*}[bt] 
\includegraphics{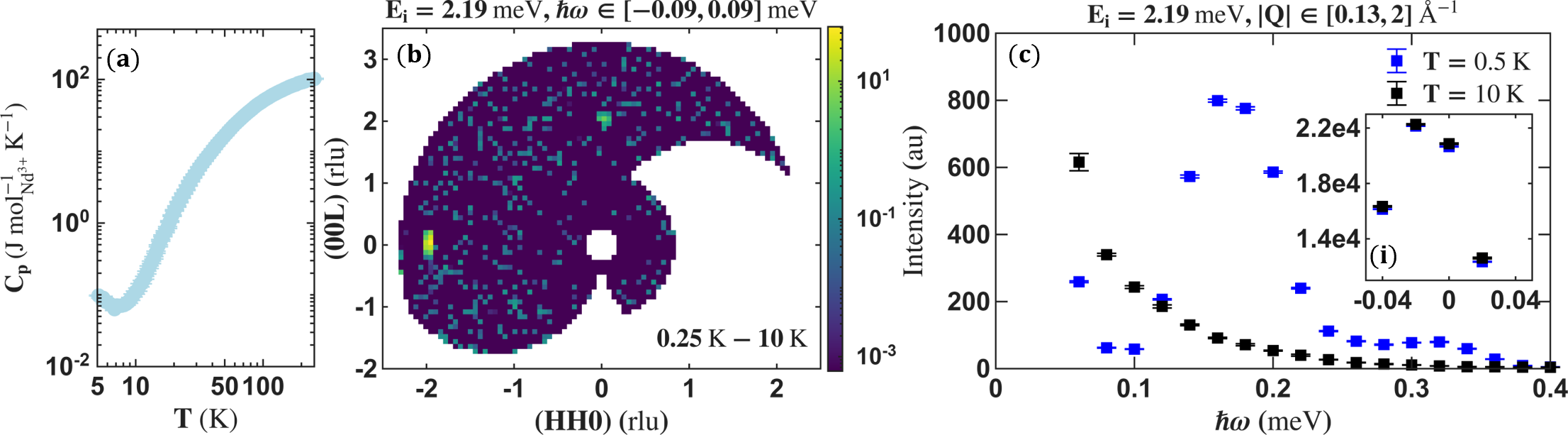}
\caption{(a) Heat capacity $C_p$ versus temperature $T$ for single-crystal \nso\ sample in the temperature range $T\in[5,250]$ K, with no background subtraction from hydrothermally grown La$_2$Sn$_2$O$_7$ crystals. (b) False-color intensity map of the $(HHL)$ plane for the co-aligned \nso\ crystal array, measured with incident energy $E_i=2.19$\,meV. The map shows the difference $I(0.25\,\mathrm{K})-I(10\,\mathrm{K})$ in the elastic channel, integrated over $[-0.09,0.09]$\,meV and within $\pm0.07$\,rlu along $[K\bar K0]$. (c) Powder-averaged intensity versus energy transfer $\hbar\omega$ at $T=0.25$ and $10$\,K for $E_i=2.19$\,meV, integrated along $[K\bar K0]$ with $K\in[-0.07,0.07]$\,rlu and over $Q\in[0.13,2]$\,\text{\AA}$^{-1}$ within the $(HHL)$ plane. Inset (i) shows the corresponding elastic intensity.}
\label{fig:SIFig3}
\end{figure*}

\begin{figure*}[tb] 
\includegraphics{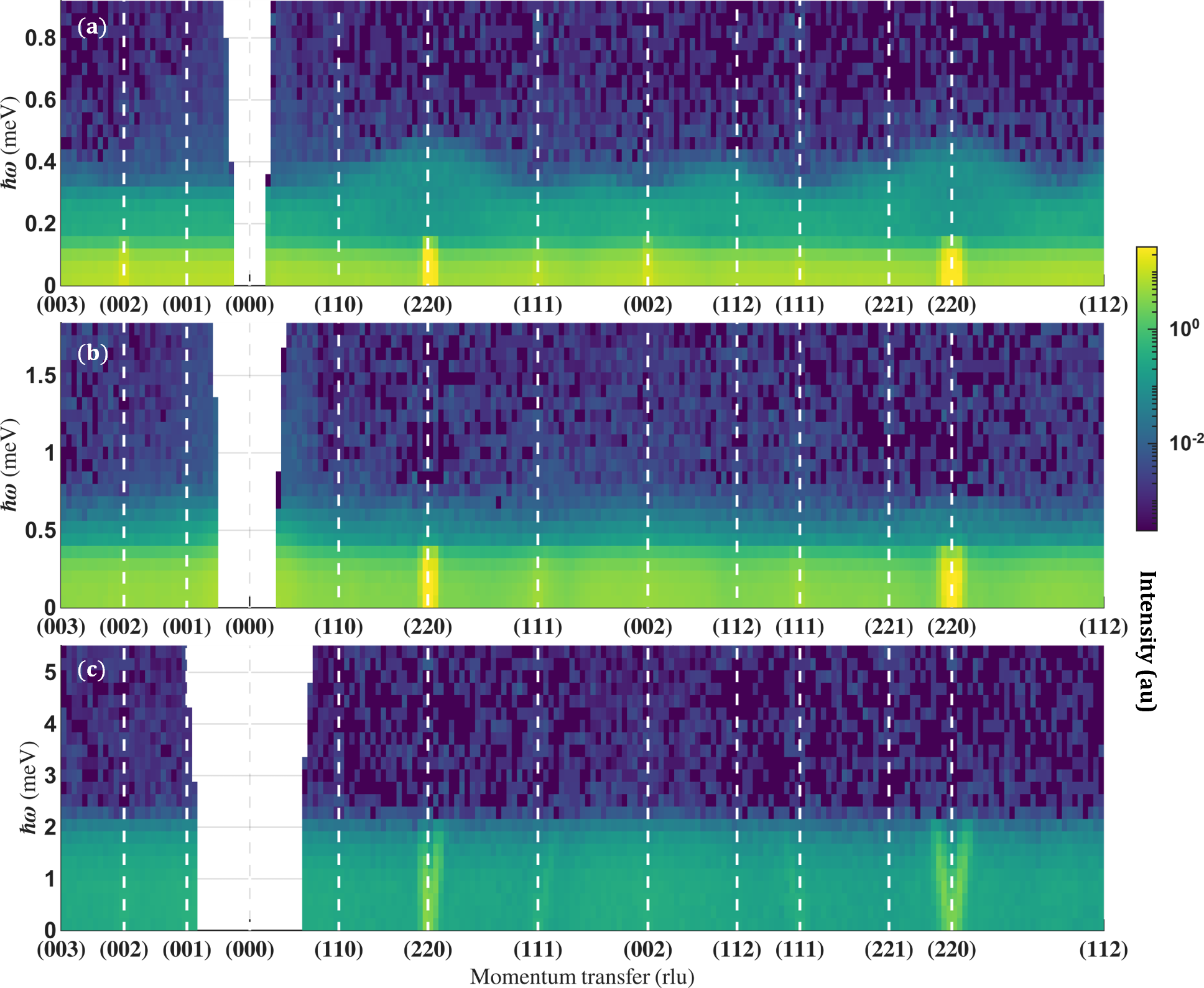}
\caption{False color map of inelastic scattering intensity $S(\mathbf{Q},\omega)$ versus energy transfer $\hbar\omega$ and momentum transfer $Q$ along a path connecting the labeled high symmetry points. Data were taken at $T=0.25$~K with incident energy (a) $E_i=3.7$ meV,(b) $7.51$ meV 
, and (c) $22.7$~meV on the LET spectrometer, integrated over a perpendicular momentum window of $\pm 0.1$~\AA$^{-1}$, and not symmetrized.}
\label{fig:SIFig4}
\end{figure*}

Single crystals of Nd$_2$Sn$_2$O$_7$ were grown by spontaneous nucleation using a high-temperature, high-pressure hydrothermal technique~\cite{powell2019hydrothermal}. Stoichiometric amounts of the corresponding dry oxides and a 5~M CsF/CsOH mineralizer solution were loaded into fine silver ampoules in a 1:4 g:mL ratio. The ampoules were weld-sealed, placed in a Tuttle-seal autoclave, and supplemented with deionized water to provide adequate counterpressure (typically $\sim$100~MPa). The sealed autoclaves were heated with ceramic band heaters to generate a cold–hot gradient of 670–680~$^\circ$C for 14–21~days, promoting crystal growth via convective transport of the nutrient solution from the hotter dissolution zone to the cooler crystallization zone. After the reaction, the autoclaves were cooled in air, and the ampoules were collected, opened, and vacuum-filtered. The recovered product was washed with deionized water, dried with acetone, and yielded well-faceted purple octahedral crystals, typically 1–20~mg in mass and millimeter-sized [Fig.~\ref{fig:SIFig1}(a--c)]. A single crystal \nso\ specimen ($\sim0.1~\mathrm{mm}^3$) was used for single-crystal X-ray diffraction (SCXRD) measurements on a Bruker Venture D8 diffractometer using Mo K$\alpha$ radiation source at room temperature, and the refined structural parameters are listed in Table~\ref{Tab:XRD}.

\begin{table*}[tb]
\centering
\caption{Atomic positions and anisotropic displacement parameters of single-crystal \nso\ collected at $T=300$~K from SCXRD measurement.}
\begin{tabular}{lccccccccc}
\hline\hline
\multicolumn{5}{l}{Lattice parameter (\AA): 10.57170(17)} & \multicolumn{5}{r}{Space group: \textit{Fd$\bar{3}$m}} \\
\hline
Atom (Wyckoff) & $x$ & $y$ & $z$ & $u_{11}$ & $u_{22}$ & $u_{33}$ & $u_{23}$ & $u_{13}$ & $u_{12}$ \\
\hline
Sn (16c) & 0.5  &0.75  &0.25 & 0.0166(8) &0.0166(8) &0.0166(8) &0.00007(10) &-0.00007(10) &-0.00007(10) \\
Nd (16d) & 0.5 & 0.5 & 0.5 & 0.0187(7) &0.0187(7) &0.0187(7) &-0.00094(8) &-0.00094(8) &-0.00094(8) \\
O (8b)   &  0.625 &0.625 &0.625 & 0.0203(17) &0.0203(17) &0.0203(17) &0 &0 &0 \\
O (48f)  & 0.4182(3) &0.625 &0.125 & 0.0187(19) &0.0196(13) &0.0196(13) &-0.0039(14) &0 &0 \\
\hline\hline
\end{tabular}\label{Tab:XRD}
\end{table*}

Heat-capacity measurements from 250 K to 1.8~K and in the dilution-refrigerator temperature range were performed using the Quantum Design heat-capacity option on a 9~T Physical Property Measurement System (PPMS). The lattice contribution was estimated from nonmagnetic, hydrothermally grown La$_2$Sn$_2$O$_7$ and subtracted from the raw data. The combined data is shown in Fig.~1(b). The slight upturn of $C_p$ below approximately 0.25~K may originate from the onset of a nuclear contribution, similar to that reported for Nd$_2$Zr$_2$O$_7$~\cite{Xu2019_AnisotropicExchange_Nd2Zr2O7}. In this regime ($T\lesssim0.25$ K), the PPMS relaxation method does not return a reliable thermal time constant, a situation that can arise when a nuclear Schottky term grows rapidly and the effective heat capacity changes during the measurement. This instability might lead to nonphysical $C_p$ values, and therefore we do not perform quantitative fitting below 0.25~K.

DC magnetic susceptibility measurements from 300-1.8 K were measured with a Magnetic Property Measurement System (MPMS3). Additional measurements from 4 K-0.4 K were performed with the Quantum Design iHe-3 He$^{3}$ insert for the MPMS3. $^4$He-range susceptibility was recorded on a $\sim$1\,mg pulverized single crystal. $^3$He-range measurements used 29.1\,mg of pulverized single-crystal \nso\ in a 100\,Oe field. The resulting datasets were normalized and combined. Magnetization versus field measurements at $T=2$~K were performed on single-crystal \nso\ samples with masses of $\sim$1.50~mg ($H \parallel [100]$), $\sim$0.93~mg ($H \parallel [111]$), and $\sim$1.36~mg ($H \parallel [110]$). The induced magnetic moments $M$ at the maximum applied field of $\mu_0H=7$~T were determined to be $1.67~\mu_\mathrm{B}/\mathrm{Nd^{3+}}$, $1.43~\mu_\mathrm{B}/\mathrm{Nd^{3+}}$, and $1.28~\mu_\mathrm{B}/\mathrm{Nd^{3+}}$ for fields along the $[100]$, $[111]$, and $[110]$ directions, respectively. For comparison, the expected saturated moments in a pyrochlore lattice with local $\langle 111 \rangle$ Ising anisotropy and effective ferromagnetic (FM) nearest-neighbor coupling are $g_JJ/\sqrt{3}$ for $[100]$, $g_JJ/2$ for $[111]$, and $g_JJ/\sqrt{6}$ for $[110]$~\cite{Fukazawa2002}. The observed ordering of induced moments (largest to smallest) and their relative ratios are consistent with these theoretical expectations for a magnetic pyrochlore with easy-axis anisotropy~\cite{Fukazawa2002}. Small quantitative deviations may originate from slight sample misalignment during mounting, uncertainties in mass due to residual impurity phases, or incomplete saturation at $\mu_0H=7$~T.

For single-crystal neutron scattering experiments, 153 individual crystals were co-aligned and affixed to one oxygen-free copper plate using GE varnish, covering a rectangular area of approximately 2.0~cm (vertical) $\times$ 1.0~cm (horizontal) on each side [Fig.~\ref{fig:SIFig1}(a)], with the $(HHL)$ plane lying in the horizontal scattering plane. The assembly behaved effectively as a single, slightly broadened crystal grain with a total mass of $\sim$0.93~g and an estimated mosaic spread of $\sim$5$^\circ$ full width at half maximum (FWHM) [Fig.~\ref{fig:SIFig1}(e--h)].  

Inelastic neutron scattering measurements were carried out on the cold neutron chopper spectrometer LET at the ISIS Neutron and Muon Source, RAL, UK. Data were collected at $T=0.25$~K and $T=10$~K using simultaneous incident energies of $E_i=2.19$, 3.7, 7.51, and 22.7~meV, enabled by the multi-chopper system. The chopper configuration was 120~Hz/120~Hz (choppers 5 and 3), 60~Hz/60~Hz (choppers 1 and 4), with chopper 2 phased to 8800 at 10~Hz. These settings provided elastic-line energy resolutions (FWHM) of approximately 0.07, 0.15, 0.40, and 2.1~meV for $E_i=2.19$, 3.7, 7.51, and 22.7~meV, respectively. For the $E_i = 2.19$ meV data shown in the main text, the accessible energy window is estimated to be $\sim 0.09$ meV based on $3\sigma$, with $\sigma = \mathrm{FWHM}/2.355$. This serves as a reasonable lower limit above which the elastic background does not dominate. The sample was rotated about the vertical $[1\bar{1}0]$ axis over a total range of $180^\circ$ in $1^\circ$ steps. Data reduction and analysis were performed using the Horace software suite~\cite{Ewings_2016_Horace}. Measurements with $E_i=22.7$~meV were specifically used to assess the sample mosaicity and alignment [Fig.~\ref{fig:SIFig1}(d--h)].  No background subtraction was applied to any of the data presented in the main text or in the Supplemental Material, except in Fig.~\ref{fig:SIFig3}(b), where it is explicitly stated that the 0.25~K\,--\,10~K difference is shown.

Figure~\ref{fig:SIFig2}(a) shows a false-color intensity map of the $(HHL)$ plane at the elastic line, collected with $E_i=3.7$~meV at $T=0.25$~K, where clear intensity is observed at the nominally forbidden (002) position. Constant-$Q$ energy cuts at several zone centers are presented in Fig.~\ref{fig:SIFig2}(b--d) for $E_i=2.19$~meV and in Fig.~\ref{fig:SIFig2}(e--g) for $E_i=3.7$~meV. At both incident energies and temperatures, finite intensity is detected at (002) [Fig.~\ref{fig:SIFig2}(b,e)]. A temperature-dependent enhancement of this signal is evident only in the $E_i=2.19$~meV data, where the intensity at $T=0.25$~K exceeds that at $T=10$~K [Fig.~\ref{fig:SIFig2}(b)], while the corresponding difference in the $E_i=3.7$~meV data is negligible [Fig.~\ref{fig:SIFig2}(e)]. The (002) reflection is also observed in measurements with $E_i=7.59$~meV (not shown), with nearly identical intensity at both $T=0.25$~K and $T=10$~K. Notably, the (002) reflection is absent in the SCXRD data collected at $T = 300$~K and $T = 100$~K. The refinement at $T=300$ K confirms a pyrochlore structure with space group $Fd\bar{3}m$ [Table~\ref{Tab:XRD}]. A genuine (002) reflection can only arise if the crystal symmetry is lowered from the pyrochlore space group $Fd\bar{3}m$, such as through a transition to the subgroup $F\bar{4}3m$~\cite{Okamoto2013_BreathingPyrochlore}. However, heat-capacity measurements on the single-crystal \nso\ sample over the temperature range [5,250]~K [Fig.~\ref{fig:SIFig3}(a)] show no indication of any structural transition. Taken together, these observations suggest that the observed (002) reflection is more likely of extrinsic origin, such as multiple scattering, rather than a genuine cooling-induced lattice distortion. Similar (002) reflections have been reported in \hto, \yto, and $\mathrm{Er_2Ti_2O_7}$, and were attributed to multiple scattering from strong allowed Bragg reflections~\cite{Baroudi2015_SymmetryLightStuffing}.

In Fig.~\ref{fig:SIFig2}(c,d,f,g), comparison of the $T=0.25$~K and $T=10$~K datasets reveals pronounced magnetic Bragg intensity at $(220)$ and $(113)$, with essentially no additional magnetic signal at $(111)$ and $(004)$. This pattern is consistent with the noncoplanar all-in–all-out (AIAO) magnetic structure reported in Ref.~\cite{Bertin_2015_Nd2Sn2O7}.

A shift of the center of the energy cuts at Bragg peaks from zero energy of up to $\sim 0.02$~meV is observed in Fig.~\ref{fig:SIFig2}(b--d) for $E_i=2.19$~meV. This behavior is most consistent with a sample-centering offset of the coaligned sample [Fig.~\ref{fig:SIFig1}(a)], as indicated by the absence of the expected scaling with $E_i$ for a constant $t_0$ offset [cf. Fig.~\ref{fig:SIFig2}(e--g) for $E_i=3.7$~meV], and by its dependence on scattering angle and detector position across different $(Q,E)$ points, which can also be observed in the $E_i=3.7$~meV data. A residual $t_0$ contribution cannot be excluded but is not expected to be dominant. For sample off-centering, the energy-scale correction is largest at the elastic line and decreases in magnitude at finite positive energy transfer for a given detector direction. Additional factors, including coarse angular sampling of sharp Bragg reflections from the coaligned crystal array, detector-dependent response, background contributions (e.g., from Cu and GE varnish), and asymmetric instrumental resolution from the chopper settings, may further distort the elastic peak shape and bias the fitted peak center, thereby complicating a unique correction of the energy scale.

\section{Additional Neutron Spectroscopy Data}
The primary data sets presented in the main text were acquired with incident energy $E_i=2.19$\,meV. Figure~\ref{fig:SIFig3}(b) shows the $(HHL)$ $Q$-map at $T=0.25$\,K in the elastic channel within resolution (FWHM $\approx 0.07$~\text{meV}), integrated over $[-0.09,0.09]$~meV, with the $T=10$\,K data subtracted as an elastic background. No residual magnetic diffuse-scattering signal is observed in the elastic channel. The corresponding powder-averaged intensity within the $(HHL)$ plane at $T=0.25$ and $10$\,K is shown in Fig.~\ref{fig:SIFig3}(c), which reveals only the spin-wave excitation discussed in the main text, confined to $\hbar\omega \in [0.1,0.4]$\,meV. 

Additional measurements at $E_i=3.7$, $7.51$, and $22.7$~meV, all at $T=0.25$~K, are shown in Fig.~\ref{fig:SIFig4}. No discernible magnetic signal is observed in the combined range $0.5$–$5$~meV across all panels. The intensity at $(220)$ for $E_i=22.7$~meV below $\hbar\omega=2$~meV is attributable to the elastic-line tail from the finite energy resolution (FWHM $\approx 2.1$~\text{meV}).

\begin{figure*}[tb] 
\includegraphics{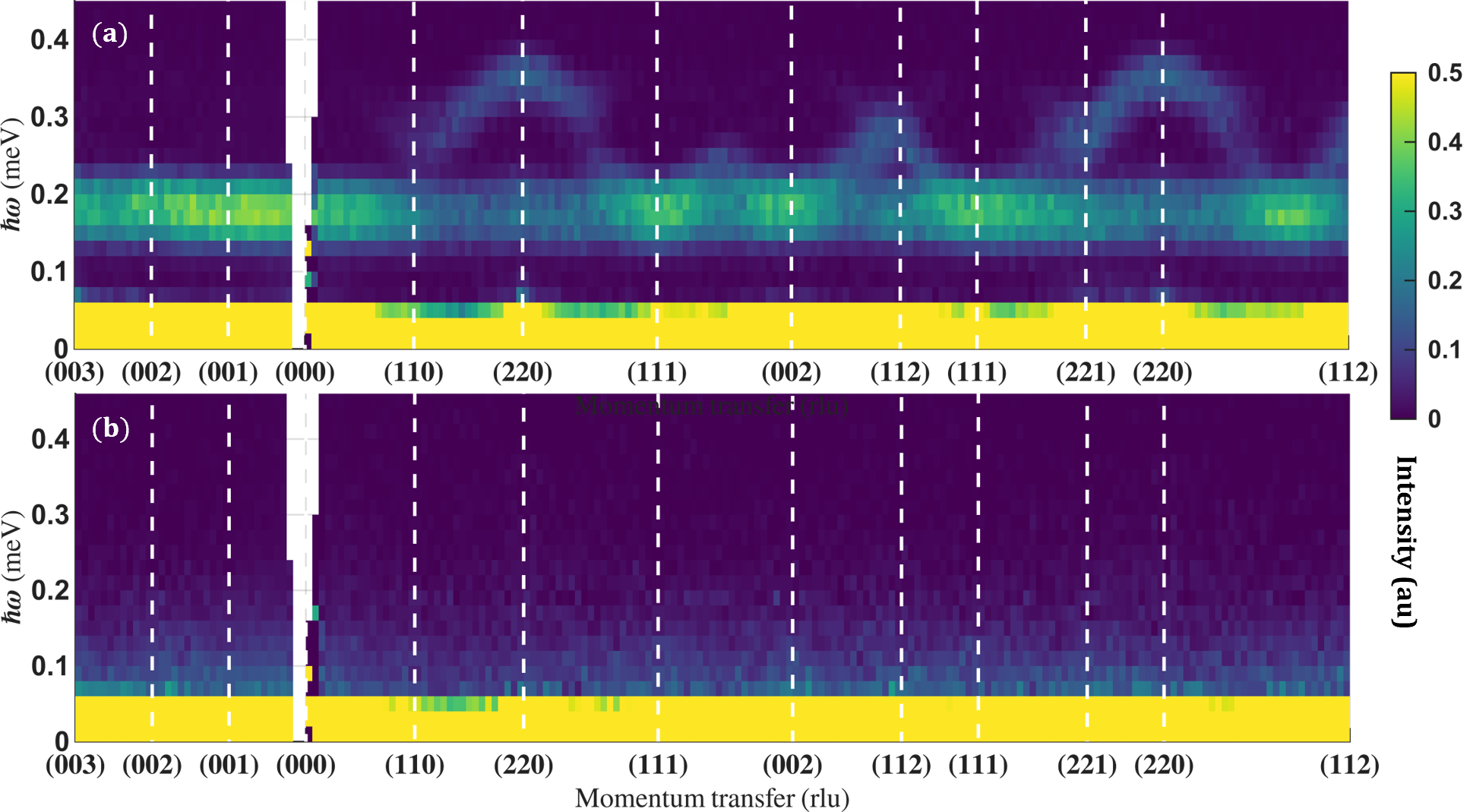}
\caption{ False color map of inelastic scattering intensity $S(\mathbf{Q},\omega)$ versus energy transfer $\hbar\omega$ and momentum transfer $Q$ along a path connecting the labeled high symmetry points. Data were taken at (a) $T=0.25$~K and (b) $T=10$~K with incident energy $E_i=2.19$~meV on the LET spectrometer, integrated over a perpendicular momentum window of $\pm 0.1$~\AA$^{-1}$, and not symmetrized.}
\label{fig:SIFig_bkg}
\end{figure*}

The $T = 10$~K neutron spectrum collected with $E_i = 2.19$\,meV is shown in Fig.~\ref{fig:SIFig_bkg}(b), alongside the $T = 0.25$~K data in Fig.~\ref{fig:SIFig_bkg}(a) (also presented in Fig.~2(a)). It is clear from Fig.~\ref{fig:SIFig_bkg}(b), Fig.~3 and Fig.~\ref{fig:SIFig5} that the spin-wave excitations 
visible at 0.25~K are entirely absent at 10~K. For this reason, no background subtraction was applied to the data presented in the main text.

\section{Details of Spin-wave model}
\subsection{LSWT Hamiltonian}
We closely follow the theoretical approach introduced in Ref.~\cite{benton2016quantum} for the isostructural sister compound \nzo. The ground state doublet of Nd$^{3+}$ in \nso\ has a dipolar-octupolar character. A natural basis choice, $\ket{\uparrow}$ and $\ket{\downarrow}$, diagonalizes the angular momentum operator $\mathbf{J}$ along the local $z$-axis, which coincides with the $C_3$ symmetry axis at each Nd$^{3+}$ site. Following Ref.~\cite{huang2014quantum}, the local axes are defined as:

\begin{align}
\begin{split}
\hat{\mathbf{z}}_1 &= \frac{1}{\sqrt{3}}(1,1,1),~~~~ ~~~\hat{\mathbf{y}}_1 = \frac{1}{\sqrt{2}}(0,1,-1),\\
\hat{\mathbf{z}}_2 &= \frac{1}{\sqrt{3}}(1,-1,-1),\quad \hat{\mathbf{y}}_2 = \frac{1}{\sqrt{2}}(-1,0,-1),\\
\hat{\mathbf{z}}_3 &= \frac{1}{\sqrt{3}}(-1,1,-1),\quad \hat{\mathbf{y}}_3 = \frac{1}{\sqrt{2}}(-1,-1,0),\\
\hat{\mathbf{z}}_4 &= \frac{1}{\sqrt{3}}(-1,-1,1),\quad \hat{\mathbf{y}}_4 = \frac{1}{\sqrt{2}}(-1,1,0),
\end{split}
\label{loc_cord}
\end{align}
and $\hat{x}_i=\hat{y}_i\times\hat{z}_i$ corresponding to the four Nd$^{3+}$ ions located at fractional coordinates $\hat{\mathbf{r}}_1 = (0.5, 0.5, 0.5)$, $\hat{\mathbf{r}}_2 = (0.5, 0.75, 0.75)$, $\hat{\mathbf{r}}_3 = (0.75, 0.5, 0.75)$, and $\hat{\mathbf{r}}_4 = (0.75, 0.75, 0.5)$ in the unit cell. Only the angular momentum component along $\hat{z}$, $\hat{J}_z$, has a non-zero matrix element within the ground state doublet, while $\hat{J}_x$ and $\hat{J}_y$ have vanishing matrix elements~\cite{Xu2015_Nd2Zr2O7_CrystalField,Lhotel2015_DipoleOctupole_Nd2Zr2O7}.

We introduce pseudospin-1/2 operators $(\tau_i^x, \tau_i^y, \tau_i^z)$ at each site $i$, such that the magnetic moment is given by $\mathbf{m}_i = g_z \uB \tau_i^z$, where $g_z$ is the effective $g$-factor along the local $\hat{z}_i$ direction. The symmetry-allowed nearest-neighbor exchange Hamiltonian is:
\begin{equation}
\begin{split}
\mathcal{H}^{\mathrm{DO}}_{\mathrm{ex}} = \sum_{\langle ij\rangle} \big[ 
& J_x \tau^x_i \tau^x_j 
+ J_y \tau^y_i \tau^y_j 
+ J_z \tau^z_i \tau^z_j \\
& + J_{xz} (\tau^x_i \tau^z_j + \tau^z_i \tau^x_j)
\big].
\end{split}
\end{equation}
where $\langle ij\rangle$ runs through all pairs of nearest neighbor Nd$^{3+}$. A global rotation of the pseudospin components $\tau_i^\alpha \rightarrow \tilde{\tau}_i^{\tilde{\alpha}}$,
\begin{align}
\begin{split}
\tilde{\tau}_i^{\tilde{x}} &= \cos(\vartheta)\, \tau_i^x + \sin(\vartheta)\, \tau_i^z, \quad 
\tilde{\tau}_i^{\tilde{y}} = \tau_i^y, \\
\tilde{\tau}_i^{\tilde{z}} &= \cos(\vartheta)\, \tau_i^z - \sin(\vartheta)\, \tau_i^x, \quad 
\tan(\vartheta) = \frac{2 J_{xz}}{J_x - J_z},
\end{split}\label{Eqn:tau_transform}
\end{align}
eliminates the $J_{xz}$ term and yields an effective XYZ model:
\begin{equation}
\mathcal{H}^{\mathrm{DO}}_{\mathrm{xyz}} = \sum_{\langle ij\rangle} \left[ 
\tilde{J}_x \tilde{\tau}_i^{\tilde{x}} \tilde{\tau}_j^{\tilde{x}} 
+ \tilde{J}_y \tilde{\tau}_i^{\tilde{y}} \tilde{\tau}_j^{\tilde{y}} 
+ \tilde{J}_z \tilde{\tau}_i^{\tilde{z}} \tilde{\tau}_j^{\tilde{z}} 
\right].\label{HXYZ}
\end{equation}

The magnetic moment at each site $i$ can then be decomposed as:
\begin{equation}
\mathbf{m}_i = g_z \uB \left[ \cos(\vartheta)\, \mathbf{m}_i^{\tilde{z}} + \sin(\vartheta)\, \mathbf{m}_i^{\tilde{x}} \right],\label{moment_def}
\end{equation}
where
\begin{equation}
\mathbf{m}_i^{\tilde{\alpha}} = \tilde{\tau}_i^{\tilde{\alpha}} \, \hat{z}_i, \qquad \alpha = x, z.\label{moment_def_com}
\end{equation}
The $\tau^y_i$ (or $\tilde{\tau}_i^{\tilde{y}}$) component transforms as a magnetic octupole and does not contribute to the dipolar magnetic moment. Following Ref.~\cite{benton2016quantum}, we identify the origin of dynamical moment fragmentation: the component $\mathbf{m}_i^{\tilde{z}}$ forms an "all-in, all-out" (AIAO) magnetic order, while fluctuations in $\mathbf{m}_i^{\tilde{x}}$ give rise to spin-wave excitations. The AIAO ground state is stabilized within the following parameter regime:
\begin{equation}
\tilde{J}_z < 0, \quad -|\tilde{J}_z| < \tilde{J}_x,\, \tilde{J}_y < 3|\tilde{J}_z|.
\end{equation}
For linear spin-wave theory (LSWT), we apply Holstein-Primakoff (HP) transformation
\begin{align}
\begin{split}
\tilde{\tau}_i^{\tilde{z}}&=S-a^{\dagger}_ia_i\\
\tilde{\tau}_i^+\equiv\tilde{\tau}_i^{\tilde{x}}+i\tilde{\tau}_i^{\tilde{y}}&=\sqrt{2S-a^{\dagger}_ia_i}~a_i\approx\sqrt{2S}~a_i\\\tilde{\tau}_i^-\equiv\tilde{\tau}_i^{\tilde{x}}-i\tilde{\tau}_i^{\tilde{y}}&=a^{\dagger}_i~\sqrt{2S-a^{\dagger}_ia_i}\approx\sqrt{2S}~a^{\dagger}_i
\end{split}\label{HP}
\end{align}
where $S=1/2$ and the HP bosons $a_i$ satisfys $[a_i,a_j^{\dagger}]=\delta_{ij}$. Substituting Eq.\ref{HP} into Eq.\ref{HXYZ} gives the LSWT Hamiltonian:
\begin{align}
\begin{split}
\mathcal{H}^{\mathrm{DO}}_{\mathrm{LSWT}}&=-3N|\tilde{J}_z|S^2+6|\tilde{J}_z|S\sum_ia^{\dagger}_ia_i\\
&+\frac{S}{2}\sum_{\langle ij \rangle}(a^{\dagger}_i,a_i)\begin{pmatrix}
\tilde{J}_x+\tilde{J}_y&\tilde{J}_x-\tilde{J}_y\\\tilde{J}_x-\tilde{J}_y&\tilde{J}_x+\tilde{J}_y
\end{pmatrix}\begin{pmatrix}a_j\\a^{\dagger}_j\end{pmatrix}
\end{split}\label{HLSWT}
\end{align}
We now perform a lattice Fourier transformation on the bosonic operators $a_i$ and $a^{\dagger}_i$, where each site $i$ is labeled by the unit cell index $n$ and one of the four sublattices $m$ within a tetrahedron:
\begin{equation}
\begin{split}
a^{\dagger}_{n,m}&=\frac{1}{\sqrt{N}}\sum_{\mathbf{k}}a^{\dagger}_{\mathbf{k}m}\exp(i\mathbf{k}\cdot(\mathbf{R}_n+\mathbf{r}_m))\\
a_{n,m}&=\frac{1}{\sqrt{N}}\sum_{\mathbf{k}}a_{\mathbf{k}m}\exp(-i\mathbf{k}\cdot(\mathbf{R}_n+\mathbf{r}_m))
\end{split}\label{boson_def}
\end{equation}

Applying these definitions to Eq.~\ref{HLSWT}, we identify the following types of terms:
\begin{widetext}
\begin{equation}
\begin{split}
\sum_{n,m} a^{\dagger}_{n,m} a_{n,m} 
&= \frac{1}{N} \sum_{\mathbf{k},\mathbf{k}',m} a^{\dagger}_{\mathbf{k}m} a_{\mathbf{k}',m} 
\exp\left[i(\mathbf{k} - \mathbf{k}') \cdot (\mathbf{R}_n + \mathbf{r}_m)\right]= \sum_{\mathbf{k}m} a^{\dagger}_{\mathbf{k}m} a_{\mathbf{k}m} \\
\\
\sum_{\langle nm,n'm' \rangle} a^{\dagger}_{n,m} a_{n',m'} 
&= \frac{1}{N} \sum_{\langle nm,n'm' \rangle} \sum_{\mathbf{k},\mathbf{k}'} 
a^{\dagger}_{\mathbf{k}m} a_{\mathbf{k}'m'} \exp\left[i\mathbf{k} \cdot (\mathbf{R}_n + \mathbf{r}_m) - i\mathbf{k}' \cdot (\mathbf{R}_{n'} + \mathbf{r}_{m'})\right]\\
&=\sum_{\mathbf{k},\langle m,m'\rangle}a^{\dagger}_{\mathbf{k}m}a_{\mathbf{k}m'}\exp\left[i\mathbf{k}\cdot\Delta\mathbf{r}_{\langle n m,n'm'\rangle}\right]\\
\sum_{\langle nm,n'm' \rangle} a_{n,m} a_{n',m'} 
&=\frac{1}{N}\sum_{\langle nm,n'm' \rangle} \sum_{\mathbf{k},\mathbf{k}'} 
a_{\mathbf{k}m} a_{\mathbf{k}'m'} \exp\left[-i\mathbf{k} \cdot (\mathbf{R}_n + \mathbf{r}_m) - i\mathbf{k}' \cdot (\mathbf{R}_{n'} + \mathbf{r}_{m'})\right]\\
&=\sum_{\mathbf{k},\langle m,m'\rangle}a_{\mathbf{k}m}a_{-\mathbf{k}m'}\exp\left[-i\mathbf{k}\cdot\Delta\mathbf{r}_{\langle n m,n'm'\rangle}\right]\\
\sum_{\langle nm,n'm' \rangle} a^{\dagger}_{n,m} a^{\dagger}_{n',m'} &=\sum_{\mathbf{k},\langle m,m'\rangle}a^{\dagger}_{\mathbf{k}m}a^{\dagger}_{-\mathbf{k}m'}\exp\left[i\mathbf{k}\cdot\Delta\mathbf{r}_{\langle n m,n'm'\rangle}\right]
\end{split}\label{Eqn:FT}
\end{equation}
\end{widetext}

Here, $\langle nm,n'm'\rangle=\langle ij\rangle$ denotes the summation over nearest-neighbor bonds connecting sublattice sites $m$ and $m'$ located in the $n$-th and $n'$-th unit cells, respectively. The notation $\langle m,m'\rangle$ refers to the same set of sublattice pairs, with $m$ assumed to belong to the reference (zeroth) unit cell. The vector $\Delta\mathbf{r}_{\langle n m,n'm'\rangle} \equiv \mathbf{R}_n + \mathbf{r}_m - \mathbf{R}_{n'} - \mathbf{r}_{m'}$ denotes the spatial displacement between the connected sublattices.

After performing the Fourier transformation, the quadratic Hamiltonian in Eq.\ref{HLSWT} takes the form \cite{smit2020magnon}
\begin{equation}
\begin{split}
\mathcal{H}_2&=\sum_{\mathbf{k}}\sum_{m,m'=1}^{4}\biggl\{A^{mm'}_{\mathbf{k}}a^{\dagger}_{\mathbf{k}m}a_{\mathbf{k}m'}\\&+\frac{1}{2}\left[B^{mm'}_{\mathbf{k}}a^{\dagger}_{\mathbf{k}m}a^{\dagger}_{-\mathbf{k}m'}+(B^{m'm}_{\mathbf{k}})^*a_{-\mathbf{k}m}a_{\mathbf{k}m'}\right]\biggr\}
\end{split}\label{H2_LWST}
\end{equation}

The hermiticity of Eq.\ref{H2_LWST} and the symmetry under relabeling $\mathbf{k}\rightarrow-\mathbf{k}$ imply that
\begin{equation}
A^{mm'}_{\mathbf{k}}=(A^{m'm}_{\mathbf{k}})^*\quad B_{\mathbf{k}}^{mm'}=B_{-\mathbf{k}}^{m'm}
\end{equation}
\subsection{Multi-flavor Bogoliubov Transformations}
The Hamiltonian in Eq.\ref{H2_LWST} can be diagonalized by multi-flavor bogoliubov transformations \cite{Colpa1978,smit2020magnon}. To proceed, we define the 4-component column vectors:
\begin{equation}
a_{\mathbf{k}}=\begin{pmatrix}a_{\mathbf{k}1}\\a_{\mathbf{k}2}\\a_{\mathbf{k}3}\\a_{\mathbf{k}4}\end{pmatrix}\quad a_{\mathbf{k}}^*=\begin{pmatrix}a^{\dagger}_{\mathbf{k}1}\\a^{\dagger}_{\mathbf{k}2}\\a^{\dagger}_{\mathbf{k}3}\\a^{\dagger}_{\mathbf{k}4}\end{pmatrix}
\end{equation}
and the adjoint row vectors
\begin{equation}
\begin{split}
a^{\dagger}_{\mathbf{k}}&=(a^{\dagger}_{\mathbf{k}1},a^{\dagger}_{\mathbf{k}2},a^{\dagger}_{\mathbf{k}3},a^{\dagger}_{\mathbf{k}4})\\
a^{T}_{\mathbf{k}}&=(a_{\mathbf{k}1},a_{\mathbf{k}2},a_{\mathbf{k}3},a_{\mathbf{k}4})
\end{split}
\end{equation}
These vectors can be combined to 8-components vectors,
\begin{equation}
\boldsymbol{\phi}_{\mathbf{k}}=\begin{pmatrix}a_{\mathbf{k}}\\a^*_{-\mathbf{k}}\end{pmatrix}=\begin{pmatrix}a_{\mathbf{k}1}\\a_{\mathbf{k}2}\\a_{\mathbf{k}3}\\a_{\mathbf{k}4}\\a^{\dagger}_{-\mathbf{k}1}\\a^{\dagger}_{-\mathbf{k}2}\\a^{\dagger}_{-\mathbf{k}3}\\a^{\dagger}_{-\mathbf{k}4}\end{pmatrix}\label{phi_def}
\end{equation}

\begin{equation}
\begin{split}
\boldsymbol{\phi}^{\dagger}_{\mathbf{k}}&=(a_{\mathbf{k}}^{\dagger},a^T_{\mathbf{-k}})\\&=(a^{\dagger}_{\mathbf{k}1},a^{\dagger}_{\mathbf{k}2},a^{\dagger}_{\mathbf{k}3},a^{\dagger}_{\mathbf{k}4},a_{-\mathbf{k}1},a_{-\mathbf{k}2},a_{-\mathbf{k}3},a_{-\mathbf{k}4})
\end{split}
\end{equation}
Eq.\ref{H2_LWST} can now be written in a compact form
\begin{equation}
\mathcal{H}_2=\frac{1}{2}\sum_{\mathbf{k}}\left[\boldsymbol{\phi}^{\dagger}_{\mathbf{k}}\mathbb{M}_{\mathbf{k}}\boldsymbol{\phi}_{\mathbf{k}}-\mathrm{Tr}\mathbf{A}_{\mathbf{k}}\right]\label{H2_LSWT_comp}
\end{equation}
where
\begin{equation}
\mathbb{M}_{\mathbf{k}}=\begin{pmatrix}\mathbf{A}_{\mathbf{k}} &\mathbf{B}_{\mathbf{k}}\\\mathbf{B}_{\mathbf{k}}^{\dagger}&\mathbf{A}^T_{-\mathbf{k}}\end{pmatrix}=\begin{pmatrix}\mathbf{A}_{\mathbf{k}} &\mathbf{B}_{\mathbf{k}}\\\mathbf{B}^*_{-\mathbf{k}}&\mathbf{A}^*_{-\mathbf{k}}\end{pmatrix},
\end{equation}
 $[\mathbf{A}_{\mathbf{k}}]^{mm'}\equiv A_{\mathbf{k}}^{mm'}$ and $[\mathbf{B}_{\mathbf{k}}]^{mm'}\equiv B_{\mathbf{k}}^{mm'}$ are $4 \times4$  block matrix satisfying $\mathbf{A}_{\mathbf{k}}=\mathbf{A}_{\mathbf{k}}^{\dagger}$ and $\mathbf{B}_{\mathbf{k}}=\mathbf{B}_{-\mathbf{k}}^{T}$.

To diagonalize Eq.~\ref{H2_LSWT_comp}, we introduce a new set of boson operators $b_{\mathbf{k}1}$, $b_{\mathbf{k}2}$, $b_{\mathbf{k}3}$, and $b_{\mathbf{k}4}$, and define the vector
\begin{equation}
\boldsymbol{\psi}_{\mathbf{k}} = \begin{pmatrix} b_{\mathbf{k}} \\ b^*_{-\mathbf{k}} \end{pmatrix} = \begin{pmatrix} b_{\mathbf{k}1} \\ b_{\mathbf{k}2} \\ b_{\mathbf{k}3} \\ b_{\mathbf{k}4} \\ b^{\dagger}_{-\mathbf{k}1} \\ b^{\dagger}_{-\mathbf{k}2} \\ b^{\dagger}_{-\mathbf{k}3} \\ b^{\dagger}_{-\mathbf{k}4} \end{pmatrix}\label{psi_def}
\end{equation}
We apply the transformation
\begin{equation}
\boldsymbol{\phi}_{\mathbf{k}} = \mathbb{T}_{\mathbf{k}} \boldsymbol{\psi}_{\mathbf{k}}\label{atransb}
\end{equation}
to Eq.~\ref{H2_LSWT_comp}, yielding
\begin{equation}
\mathcal{H}_2 = \frac{1}{2} \sum_{\mathbf{k}} \left[ \boldsymbol{\psi}^{\dagger}_{\mathbf{k}} \mathbb{T}^{\dagger}_{\mathbf{k}} \mathbb{M}_{\mathbf{k}} \mathbb{T}_{\mathbf{k}} \boldsymbol{\psi}_{\mathbf{k}} - \mathrm{Tr} \, \mathbf{A}_{\mathbf{k}} \right]\label{Hquad_inb}
\end{equation}
The transformation matrix $\mathbb{T}_{\mathbf{k}}$ is an $8 \times 8$ matrix that must satisfy the following conditions:

\paragraph{Diagonalization:} The Hamiltonian is diagonalized by $\mathbb{T}_{\mathbf{k}}$:
\begin{equation}
\mathbb{D}_{\mathbf{k}} = \mathbb{T}^{\dagger}_{\mathbf{k}} \mathbb{M}_{\mathbf{k}} \mathbb{T}_{\mathbf{k}}
\end{equation}
where $\mathbb{D}_{\mathbf{k}}$ is diagonal with eigenvalues $d_{\mathbf{k}i}$.

\paragraph{Bosonic Commutation Relations:} The transformed operators must satisfy canonical boson commutation relations:
\begin{equation}
\mathbb{T}^{\dagger}_{\mathbf{k}} \mathbb{G} \mathbb{T}_{\mathbf{k}} = \mathbb{G} = \mathbb{T}_{\mathbf{k}} \mathbb{G} \mathbb{T}^{\dagger}_{\mathbf{k}}
\end{equation}
with the metric matrix
\begin{equation}
\mathbb{G} = \begin{pmatrix} \mathbf{1}_{4\times4} & 0 \\ 0 & -\mathbf{1}_{4\times4} \end{pmatrix}
\end{equation}
Thus, $\mathbb{T}_{\mathbf{k}}$ is a symplectic matrix, not unitary.

\paragraph{Permutation Condition:} Define the permutation matrix:
\begin{equation}
\mathbb{P} = \begin{pmatrix} 0 & \mathbf{1}_{4\times4} \\ \mathbf{1}_{4\times4} & 0 \end{pmatrix}
\end{equation}
We require:
\begin{align}
\boldsymbol{\phi}^*_{-\mathbf{k}} &= \mathbb{P} \boldsymbol{\phi}_{\mathbf{k}} \\
\boldsymbol{\psi}^*_{-\mathbf{k}} &= \mathbb{P} \boldsymbol{\psi}_{\mathbf{k}}
\end{align}
which implies:
\begin{equation}
\mathbb{P} \mathbb{T}_{\mathbf{k}} = \mathbb{T}^*_{-\mathbf{k}} \mathbb{P}, \quad \mathbb{P} \mathbb{T}_{\mathbf{k}} \mathbb{P} = \mathbb{T}^*_{-\mathbf{k}} \label{Tk-k}
\end{equation}
Hence, $\mathbb{T}_{-\mathbf{k}}$ can be derived from $\mathbb{T}_{\mathbf{k}}$, and has a block structure:
\begin{equation}
\mathbb{T}_{\mathbf{k}} = \begin{pmatrix} \mathbf{Q}_{\mathbf{k}} & \mathbf{R}_{\mathbf{k}} \\ \mathbf{R}^*_{-\mathbf{k}} & \mathbf{Q}^*_{-\mathbf{k}} \end{pmatrix}
\end{equation}
with $\mathbf{Q}_{\mathbf{k}}$ and $\mathbf{R}_{\mathbf{k}}$ as independent $4 \times 4$ matrices.

To numerically obtain the Bogoliubov transformation, we define:
\begin{equation}
\mathbb{T}_{\mathbf{k}} = (\boldsymbol{v}_{\mathbf{k}1}, \boldsymbol{v}_{\mathbf{k}2}, \ldots, \boldsymbol{v}_{\mathbf{k}8}) \label{Tv_def}
\end{equation}
Conditions (1) and (2) become:
\begin{align}
\boldsymbol{v}^{\dagger}_{\mathbf{k}i} \mathbb{M}_{\mathbf{k}} \boldsymbol{v}_{\mathbf{k}j} &= \delta_{ij} d_{\mathbf{k}i} \label{vdiag} \\
\boldsymbol{v}^{\dagger}_{\mathbf{k}i} \mathbb{G} \boldsymbol{v}_{\mathbf{k}j} &= g_{ij} \label{vnorm}
\end{align}
with $g_{ij} = \delta_{ij}$ for $i \leq 4$ and $g_{ij} = -\delta_{ij}$ for $i > 4$.

This leads to the generalized eigenvalue problem:
\begin{equation}
\mathbb{M}_{\mathbf{k}}\boldsymbol{v}_{\mathbf{k}}=\omega\mathbb{G}\boldsymbol{v}_{\mathbf{k}}
\end{equation}
which is equivalent to
\begin{equation}
\mathbb{M}^{\mathrm{dyn}}_{\mathbf{k}}\boldsymbol{v}_{\mathbf{k}}=\omega\boldsymbol{v}_{\mathbf{k}}\label{dyn_eigen}
\end{equation}
here we define $\mathbb{M}^{\mathrm{dyn}}_{\mathbf{k}}$ which is non-Hermitian. Suppose we find 8 linearly independent eigenvectors $\{\boldsymbol{v}_{\mathbf{k}i}\}$ satisfying Eq.~\ref{dyn_eigen} and normalize them to obey Eq.~\ref{vnorm}, then by construction we have
\begin{equation}
\boldsymbol{v}^{\dagger}_{\mathbf{k}i} \mathbb{M}_{\mathbf{k}} \boldsymbol{v}_{\mathbf{k}j} = \omega_{\mathbf{k}j} \boldsymbol{v}^{\dagger}_{\mathbf{k}i} \mathbb{G} \boldsymbol{v}_{\mathbf{k}j} = \delta_{ij} \omega_{\mathbf{k}i} g_{ii}
\end{equation}
implying $d_{\mathbf{k}i} = \omega_{\mathbf{k}i} g_{ii}$. $\omega_{\mathbf{k}i}$ is therefore real and corresponds to $d_{\mathbf{k}i}$ up to a possible sign change.

From the Hermiticity of $\mathbb{M}_{\mathbf{k}}$, we find:
\begin{equation}
0 = (\mathbb{M}_{\mathbf{k}} \boldsymbol{v}_{\mathbf{k}i})^{\dagger} \boldsymbol{v}_{\mathbf{k}j} - \boldsymbol{v}^{\dagger}_{\mathbf{k}i} \mathbb{M}_{\mathbf{k}} \boldsymbol{v}_{\mathbf{k}j} = (\omega_{\mathbf{k}i} - \omega_{\mathbf{k}j}) \boldsymbol{v}^{\dagger}_{\mathbf{k}i} \mathbb{G} \boldsymbol{v}_{\mathbf{k}j}
\end{equation}
where $\boldsymbol{v}_{\mathbf{k}i}$ and $\boldsymbol{v}_{\mathbf{k}j}$ are two linearly independent eigenvectors ($i\neq j$). This procedure ensures that the bosonic commutation condition [Eq.~\ref{vnorm}] is satisfied when $\omega_{\mathbf{k}i} \neq \omega_{\mathbf{k}j}$. In the case of a $m$-th fold degenerate eigenvalue, it is necessary to apply a generalized Gram-Schmidt orthogonalization:

\begin{equation}
\boldsymbol{v}_{\mathbf{k}i} \rightarrow \boldsymbol{v}_{\mathbf{k}i} - \sum_{j=i+1}^{m} \boldsymbol{v}_{\mathbf{k}j} \frac{\boldsymbol{v}^{\dagger}_{\mathbf{k}j} \mathbb{G} \boldsymbol{v}_{\mathbf{k}i}}{\boldsymbol{v}^{\dagger}_{\mathbf{k}j} \mathbb{G} \boldsymbol{v}_{\mathbf{k}j}}, \quad i = 1, \ldots, m-1,
\end{equation}
to enforce the generalized orthogonality condition $\boldsymbol{v}_{\mathbf{k}j}^{\dagger} \mathbb{G} \boldsymbol{v}_{\mathbf{k}i} = 0$ for vectors $i \neq j$ corresponding to degenerate values of $\omega_{\mathbf{k}}$.

In practice, we begin by diagonalizing the dynamical matrix $\mathbb{M}^{\mathrm{dyn}}_{\mathbf{k}}$:
\begin{align}
\tilde{\mathbb{T}}_{\mathbf{k}} &= \left( \tilde{\boldsymbol{v}}_{\mathbf{k}1}, \tilde{\boldsymbol{v}}_{\mathbf{k}2}, \ldots, \tilde{\boldsymbol{v}}_{\mathbf{k}8} \right), \\
\tilde{\mathbb{D}}_{\mathbf{k}} &= \tilde{\mathbb{T}}_{\mathbf{k}}^{-1} \mathbb{M}^{\mathrm{dyn}}_{\mathbf{k}} \tilde{\mathbb{T}}_{\mathbf{k}}.
\end{align}

We then normalize and orthogonalize the eigenvectors $\{ \tilde{\boldsymbol{v}}_{\mathbf{k}i} \}$ in two steps. First, we normalize each vector as:
\begin{equation}
\tilde{\boldsymbol{v}}_{\mathbf{k}i} \rightarrow \frac{\tilde{\boldsymbol{v}}_{\mathbf{k}i}}{\left( \tilde{\boldsymbol{v}}^{\dagger}_{\mathbf{k}i} \mathbb{G} \tilde{\boldsymbol{v}}_{\mathbf{k}i} \right)}, \quad i = 1, \ldots, 8.
\end{equation}

Next, for each degenerate subspace of dimension $m$, consisting of eigenvectors $\{ \tilde{\boldsymbol{v}}_{\mathbf{k}n}, \tilde{\boldsymbol{v}}_{\mathbf{k}n+1}, \ldots, \tilde{\boldsymbol{v}}_{\mathbf{k}n+m} \}$, we apply a generalized Gram-Schmidt orthogonalization:
\begin{equation}
\tilde{\boldsymbol{v}}_{\mathbf{k}i} \rightarrow \tilde{\boldsymbol{v}}_{\mathbf{k}i} - \sum_{j=i+1}^{n+m} \tilde{\boldsymbol{v}}_{\mathbf{k}j} \frac{ \tilde{\boldsymbol{v}}^{\dagger}_{\mathbf{k}j} \mathbb{G} \tilde{\boldsymbol{v}}_{\mathbf{k}i} }{ \tilde{\boldsymbol{v}}^{\dagger}_{\mathbf{k}j} \mathbb{G} \tilde{\boldsymbol{v}}_{\mathbf{k}j} }, \quad i = n, \ldots, n+m-1.
\end{equation}

This iterative process yields a normalized and orthogonal set of eigenvectors $\{ \boldsymbol{v}_{\mathbf{k}i} \}$ that satisfy Eqs.~\ref{vdiag} and~\ref{vnorm}, thereby enabling the construction of $\mathbb{T}_{\mathbf{k}}$ and $\mathbb{T}_{-\mathbf{k}}$ using Eqs.~\ref{Tv_def} and \ref{Tk-k}. Substituting Eq.\ref{atransb} to Eq.\ref{Hquad_inb} gives
\begin{equation}
\mathcal{H}_2=\sum_{\mathbf{k}}\sum_{\lambda=1}^4\omega_{\mathbf{k}\lambda}\left(b^{\dagger}_{\mathbf{k}\lambda}b_{\mathbf{k}\lambda}+\frac{1}{2}\right)\label{Eqn:norm_Bogo}
\end{equation}
The four bands of magnon consist of two degenerate flat modes and two dispersive modes.

\begin{figure}[t] 
\includegraphics{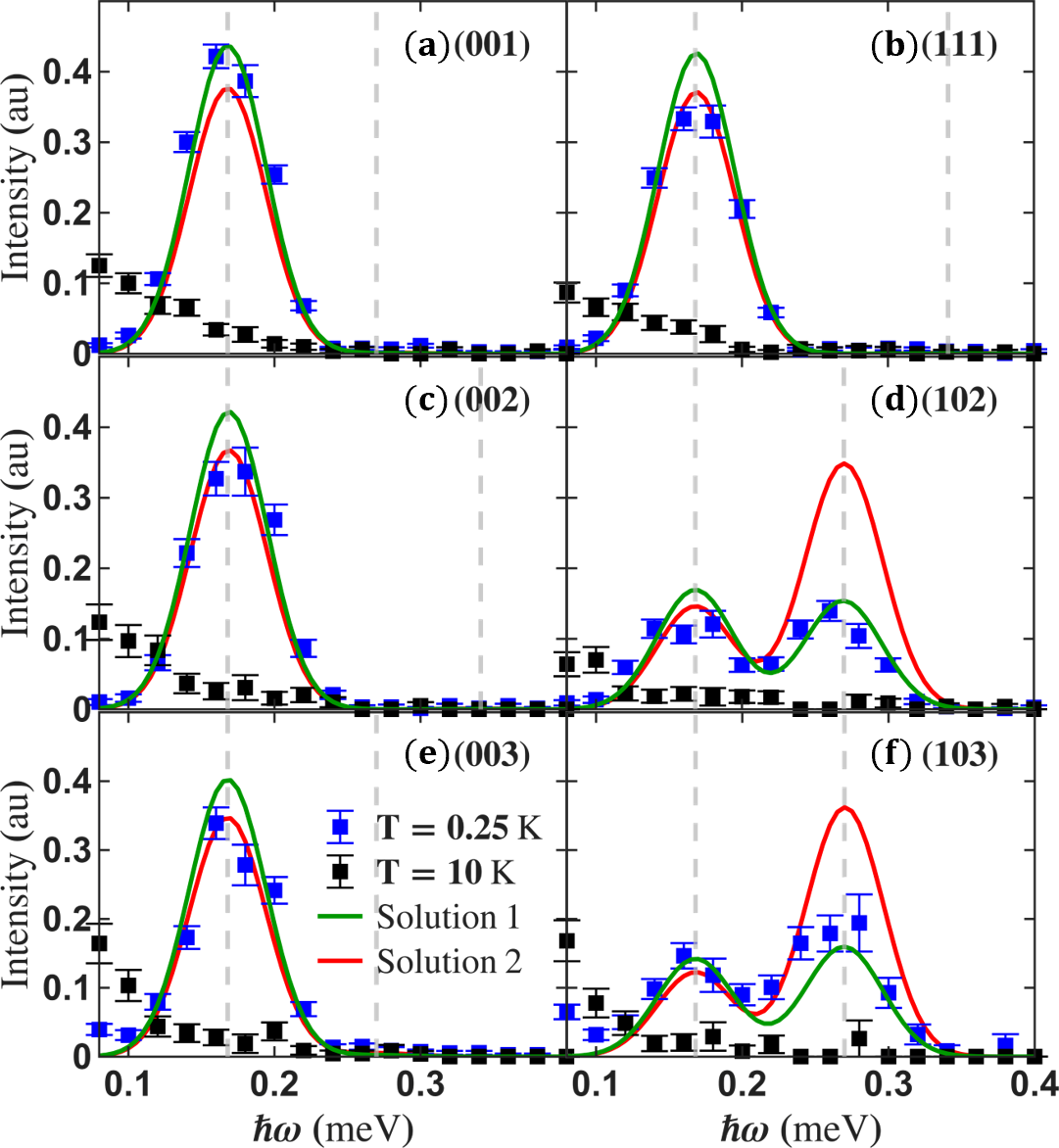}
\caption{Additional data. (a–f) Constant-$Q$ energy cuts at the indicated positions, integrated over $[-0.1,0.1]$~rlu along each principal reciprocal-space direction and measured with $E_i=2.19$~meV. Blue/black curves: $T=0.25$~K/$10$~K. Green/red curves: simulations for Solution~1 (Eq.~\ref{sol1})/Solution~2 (Eq.~\ref{sol2}), averaged over the same $Q$-windows as the data and convolved with the LET instrumental energy resolution at $E_i=2.19$~meV and scaled by the same independent global factors used in Fig.~3 of the main text, so that the combined simulated intensity across (a–f) and Fig.~3(a–d) matches experiment. Data in (a,b,c,e) are averaged over symmetry-equivalent $\mathbf{Q}$ points in the $(HHL)$ plane using the $Fd\bar{3}m$ space-group symmetry; (d,f) are not symmetrized. Gray dashed lines mark the calculated spin-wave energies.}
\label{fig:SIFig5}
\end{figure}

\subsection{Calculation of Magnetic Structure Factor}
The structure factor for magnetic neutron scattering is
\begin{equation}
\begin{split}
\mathcal{S}(\mathbf{k},\omega)&=\frac{1}{2\pi}\int_{-\infty}^{\infty}dt\exp(-i\omega t)\sum_{\mu\nu}(\delta_{\mu\nu}-\frac{k_\mu k_\nu}{k^2})\\&\times\langle m^{\mu}(-\mathbf{k},0) m^{\nu}(\mathbf{k},t)\rangle
\end{split}\label{neut_cross}
\end{equation}
Here $\mu$, $\nu$ labels the global $\hat{x}$, $\hat{y}$, $\hat{z}$ directions. We focus on the correlation function:
\begin{equation}
\begin{split}
&\langle m^{\mu}(-\mathbf{k},0) m^{\nu}(\mathbf{k},t)\rangle\\=&\sum_{\alpha,\beta=1}^4\langle m_{\alpha}(-\mathbf{k},0)m_{\beta}(\mathbf{k},t)(\hat{z}_{\alpha}\cdot\hat{\mu})(\hat{z}_{\beta}\cdot\hat{\nu})\rangle
\end{split}
\end{equation}
Here, $m_{\alpha}$ and $m_{\beta}$ are the magnetic moments on sublattices $\alpha$ and $\beta$, respectively. Following Eqs.~\ref{moment_def},\ref{moment_def_com}, we have the real-space expression
\begin{equation}
m_{n,\alpha}=g_z\uB (\cos(\vartheta)\tilde{\tau}^{\tilde{z}}_{n,\alpha}+\sin(\vartheta)\tilde{\tau}^{\tilde{x}}_{n,\alpha})
\end{equation}
where $n$ labels the unit cell. We note that correlation functions between the $\tilde{\tau}^{\tilde{z}}$ components give rise to static long-range correlations, which result in magnetic Bragg peaks associated with the AIAO phase. In contrast, correlations between the $\tilde{\tau}^{\tilde{x}}$ components generate dynamical spin-wave excitations. Focus on the spin-wave excitation, we have (omitting the time label for now)
\begin{equation}
\begin{split}
m^{\tilde{x}}_{\beta}(\mathbf{k})=&\frac{1}{\sqrt{N}}\sum_n\exp(-i\mathbf{k}\cdot(\mathbf{R_n+\mathbf{r}_{\beta}}))\\\times&\frac{\sqrt{2S}}{2}(a_{n,\beta}+a_{n,\beta}^{\dagger})\cdot(g_z\uB\sin(\vartheta))
\end{split}
\end{equation}
Substituting the definition in Eq.\ref{boson_def} into the above equations gives
\begin{equation}
\begin{split}
m^{\tilde{x}}_{\beta}(\mathbf{k})=&\frac{\sqrt{2S}}{2}g_z\uB\sin(\vartheta)(a^{\dagger}_{\mathbf{k}\beta}+a_{-\mathbf{k}\beta}).
\end{split}
\end{equation}
Similarly,
\begin{equation}
\begin{split}
m^{\tilde{x}}_{\alpha}(-\mathbf{k})=&\frac{\sqrt{2S}}{2}g_z\uB\sin(\vartheta)(a^{\dagger}_{-\mathbf{k}\alpha}+a_{\mathbf{k}\alpha})
\end{split}
\end{equation}
With the definition in Eq.\ref{phi_def}, we have
\begin{equation}
\begin{split}
&\langle m_\alpha^{\tilde{x}}(-\mathbf{k},0) m_\beta^{\tilde{x}}(\mathbf{k},t)\rangle\\=&\frac{S}{2}(g_z\uB\sin(\vartheta))^2\biggl\langle(\boldsymbol{\phi}_{\mathbf{k}\alpha}(0)+\boldsymbol{\phi}_{\mathbf{k}\alpha+4}(0))\\\times&(\boldsymbol{\phi}_{-\mathbf{k}\beta}(t)+\boldsymbol{\phi}_{-\mathbf{k}\beta+4}(t))\biggr\rangle\\&~
\end{split}
\end{equation}
Substituting the Bogoliubov transformation in Eq.\ref{psi_def},\ref{atransb} gives
\begin{equation}
\begin{split}
&\langle m_\alpha^{\tilde{x}}(-\mathbf{k},0) m_\beta^{\tilde{x}}(\mathbf{k},t)\rangle\\=&\frac{S}{2}(g_z\uB\sin(\vartheta))^2\\\times&\biggl\langle\sum_{\delta=1}^8\left(\mathbb{T}^{\alpha,\delta}_{\mathbf{k}}\boldsymbol{\psi}_{\mathbf{k}\delta}(0)+\mathbb{T}^{\alpha+4,\delta}_{\mathbf{k}}\boldsymbol{\psi}_{\mathbf{k}\delta}(0)\right)\\\times&\sum_{\delta'=1}^8\left(\mathbb{T}^{\beta,\delta'}_{-\mathbf{k}}\boldsymbol{\psi}_{-\mathbf{k}\delta'}(t)+\mathbb{T}^{\beta+4,\delta'}_{-\mathbf{k}}\boldsymbol{\psi}_{-\mathbf{k}\delta'}(t)\right)\biggr\rangle
\end{split}\label{ab_corr}
\end{equation}
The results of Eq.\ref{ab_corr} consist of a polynomial of $\langle b_{\mathbf{k}\lambda}^{(\dagger)}b_{\mathbf{k}'\lambda'}^{(\dagger)}\rangle$, here $\lambda,\lambda' = 1,2,3,4$ label the normal modes defined in Eq.~\ref{Eqn:norm_Bogo}. At finite temperatures, the relevant bosonic correlation functions are
\begin{equation}
\begin{aligned}
\frac{1}{2\pi}\int_{-\infty}^{\infty} dt\, e^{-i\omega t}
\langle b_{\mathbf{k}\lambda}(0)b^{\dagger}_{\mathbf{k}\lambda}(t) \rangle
&= (1 + n_{\mathrm{B}})\, \delta(\omega - \omega_{\mathbf{k}\lambda}), \\
\frac{1}{2\pi}\int_{-\infty}^{\infty} dt\, e^{-i\omega t}
\langle b^{\dagger}_{\mathbf{k}\lambda}(0)b_{\mathbf{k}\lambda}(t)\rangle
&= n_{\mathrm{B}}\, \delta(\omega + \omega_{\mathbf{k}\lambda}),
\end{aligned}\label{Eqn:Bose_corr}
\end{equation}
where \(n_{\mathrm{B}} \equiv n_{\mathrm{B}}(\omega_{\mathbf{k}\lambda})\) is the Bose occupation factor, with
\(n_{\mathrm{B}}(\omega)=\bigl[\exp(\hbar\omega/k_{\mathrm B}T)-1\bigr]^{-1}\). Notice that for low temperature below the gap energy $\Delta_1$ of the flat modes $(k_{\mathrm{B}}T\ll\Delta_1)$, we mostly need to consider the ground state (vaccum of $b_{\lambda}$) in the thermal average $\langle...\rangle$.

The neutron cross section on the energy-loss side ($\hbar\omega > 0$) can be obtained from Eq.~\ref{ab_corr} by retaining terms with $\delta=\lambda$ and $\delta'=\lambda+4$, yielding
\begin{equation}
\begin{split}
&\langle m_\alpha^{\tilde{x}}(-\mathbf{k},0) m_\beta^{\tilde{x}}(\mathbf{k},t)\rangle\\=&\frac{S}{2}(g_z\uB\sin(\vartheta))^2\\\times&\sum_{\lambda=1}^4(\mathbb{T}^{\alpha,\lambda}_{\mathbf{k}}\mathbb{T}^{\beta,\lambda+4}_{-\mathbf{k}}+\mathbb{T}^{\alpha,\lambda}_{\mathbf{k}}\mathbb{T}^{\beta+4,\lambda+4}_{-\mathbf{k}}\\+&\mathbb{T}^{\alpha+4,\lambda}_{\mathbf{k}}\mathbb{T}^{\beta,\lambda+4}_{-\mathbf{k}}+\mathbb{T}^{\alpha+4,\lambda}_{\mathbf{k}}\mathbb{T}^{\beta+4,\lambda+4}_{-\mathbf{k}})\langle b_{\mathbf{k}\lambda}(0)b^{\dagger}_{\mathbf{k}\lambda}(t)\rangle
\end{split}
\end{equation}
where we retain only the energy-loss (multipolar magnon creation) contribution. Similarly, we could derive cross section on the energy-gain side ($\hbar\omega < 0$) by the detailed-balance equation $S(-\mathbf{k},-\omega)=\exp(-\hbar\omega/k_\mathrm{B}T)S(\mathbf{k},\omega)$. 

Substituting these into Eq.~\ref{neut_cross} yields the dynamical magnetic neutron structure factor:
\begin{equation}
\begin{split}
\mathcal{S}^{\tilde{x}}(\mathbf{k},\omega) &= \frac{S}{2}(g_z \mu_\mathrm{B} \sin\vartheta)^2 \sum_{\mu,\nu} \left( \delta_{\mu\nu} - \frac{k_{\mu}k_{\nu}}{k^2} \right) \\
&\times \sum_{\lambda=1}^{4} s_{\lambda}(\mathbf{k}) \left[ (1 + n_{\mathrm{B}}(\omega_{\mathbf{k}\lambda}))\, \delta(\omega - \omega_{\mathbf{k}\lambda}) \right. \\
&\qquad\qquad\qquad\left. +\, n_{\mathrm{B}}(\omega_{\mathbf{k}\lambda})\, \delta(\omega + \omega_{\mathbf{k}\lambda}) \right]
\end{split}\label{Eqn:structure_factor_pn}
\end{equation}
where we define the coefficients for the four modes as
\begin{equation}
\begin{split}
 &s_{\lambda}(\mathbf{k})=\sum_{\alpha,\beta=1}^4(\hat{z}_{\alpha}\cdot\hat{\mu})(\hat{z}_{\beta}\cdot\hat{\nu})\\\times&(\mathbb{T}^{\alpha,\lambda}_{\mathbf{k}}\mathbb{T}^{\beta,\lambda+4}_{-\mathbf{k}}+\mathbb{T}^{\alpha,\lambda}_{\mathbf{k}}\mathbb{T}^{\beta+4,\lambda+4}_{-\mathbf{k}}\\+&\mathbb{T}^{\alpha+4,\lambda}_{\mathbf{k}}\mathbb{T}^{\beta,\lambda+4}_{-\mathbf{k}}+\mathbb{T}^{\alpha+4,\lambda}_{\mathbf{k}}\mathbb{T}^{\beta+4,\lambda+4}_{-\mathbf{k}})
\end{split}
\end{equation}
For comparison with the experimentally measured neutron scattering cross section,
\begin{equation}
\frac{d^2\sigma}{d\Omega\, dE_f}(\mathbf{k},\omega) = \frac{k_f}{k_i} (\gamma r_0)^2 |f(\mathbf{k})|^2 \mathcal{S}^{\tilde{x}}(\mathbf{k},\omega), \label{cross_section}
\end{equation}
we multiply the calculated dynamical magnetic structure factor $\mathcal{S}^{\tilde{x}}(\mathbf{k},\omega)$ by the squared magnetic form factor $|f(\mathbf{k})|^2$ of Nd$^{3+}$ ions. In Eq.~\ref{cross_section}, $r_0 = 2.818 \times 10^{-15}$~m is the classical electron radius, and $\gamma = -1.913$ is the magnetic dipole moment of the neutron in units of the nuclear Bohr magneton. The quantities $k_i$ and $k_f$ denote the incident and scattered neutron momenta, respectively. These prefactors do not affect the $\mathbf{k}$-dependence of the observed inelastic signal. 

\subsection{Evaluation of Rotation Angle \texorpdfstring{$\vartheta$}{theta}}

To estimate the fraction of ordered moment in the ground state relative to the total moment of Nd$^{3+}$ in this model, we calculate~\cite{benton2016quantum}
\begin{equation}
\langle\tilde{\tau}^{\tilde{z}}_i\rangle = S - \langle a^{\dagger}_{n,m}a_{n,m}\rangle,
\end{equation}
where site $i$ is again specified by the unit cell $n$ and one of the four sublattices $m$.  
Using Eqs.~\ref{Eqn:FT}, \ref{phi_def}, and \ref{atransb}, we obtain
\begin{align}
\langle a^{\dagger}_{n,m}a_{n,m}\rangle
&= \frac{1}{N}\sum_{\mathbf{k}}\langle a^{\dagger}_{\mathbf{k}m}a_{\mathbf{k}m}\rangle \nonumber\\
&= \frac{1}{N}\sum_{\mathbf{k}}\langle \boldsymbol{\phi}_{-\mathbf{k},m+4}\,\boldsymbol{\phi}_{\mathbf{k},m}\rangle \label{Eqn:occ} \\
&= \frac{1}{N}\sum_{\mathbf{k}}\sum_{\delta,\delta'=1}^8
\langle \mathbb{T}_{-\mathbf{k}}^{m+4,\delta}\,\boldsymbol{\psi}_{-\mathbf{k},\delta}\,
\mathbb{T}_{\mathbf{k}}^{m,\delta'}\,\boldsymbol{\psi}_{\mathbf{k},\delta'}\rangle. \nonumber
\end{align}
Similar to Eq.\ref{Eqn:Bose_corr}, the relevant bosonic correlation functions are $\langle b_{-\mathbf{k}\lambda}b^{\dagger}_{-\mathbf{k}\lambda}\rangle = 1+n_{\mathrm{B}}(\omega_{-\mathbf{k}\lambda})$, arising from  $\delta=\lambda$, $\delta'=\lambda+4$, and $\langle b^{\dagger}_{\mathbf{k}\lambda}b_{\mathbf{k}\lambda}\rangle = n_{\mathrm{B}}(\omega_{\mathbf{k}\lambda})$, arising from  $\delta=\lambda+4$, $\delta'=\lambda$, yielding
\begin{equation}
\begin{split}
\langle a^{\dagger}_{n,m}a_{n,m}\rangle
&= \frac{1}{N}\sum_{\mathbf{k}}\sum_{\lambda=1}^4
\left(\mathbb{T}_{-\mathbf{k}}^{m+4,\lambda}\,\mathbb{T}_{\mathbf{k}}^{m,\lambda+4}(1+n_{\mathrm{B}}(\omega_{-\mathbf{k}\lambda}))\right.\\&+\left.\mathbb{T}_{-\mathbf{k}}^{m+4,\lambda+4}\,\mathbb{T}_{\mathbf{k}}^{m,\lambda}n_{\mathrm{B}}(\omega_{\mathbf{k}\lambda})\right)\\&\approx\frac{1}{N}\sum_{\mathbf{k}}\sum_{\lambda=1}^4
\mathbb{T}_{-\mathbf{k}}^{m+4,\lambda}\,\mathbb{T}_{\mathbf{k}}^{m,\lambda+4},\quad k_{\mathrm{B}}T\ll\Delta_1.\label{Eqn:aN_sum}
\end{split}
\end{equation}
In practice, we use a $10\times10\times10$ array of conventional cubic unit cells (along $\hat{x}$, $\hat{y}$, and $\hat{z}$), sum over $4{,}000$ $\mathbf{k}$ points in the first Brillouin zone, and set $N=4{,}000$ in Eq.~(\ref{Eqn:aN_sum}). The fraction of ordered moment is then given by
\begin{equation}
\frac{m_{\mathrm{ord}}}{m_{\mathrm{sat}}}
= \cos\vartheta\left(\frac{S - \langle a^{\dagger}_{n,m}a_{n,m}\rangle}{S}\right).\label{Eqn:cal_m_ratio}
\end{equation}
\subsection{Calculation of Heat Capacity from Spin-wave Excitations}
Using the interaction parameters and linear spin-wave spectra obtained in Sec.\ref{Sec:Determin}, we numerically evaluate the single-particle density of states $g(E)$ (normalized per Nd ion such that $\int_{0}^{\infty} g(E)\, dE = 1$) and compute the spin-wave contribution to the specific heat $C_{\mathrm{SW}}$ following the standard bosonic formalism~\cite{kang2023phonon}:
\begin{equation}
\begin{split}
C_{\mathrm{SW}}(T) &= \int_{0}^{\infty} dE\, g(E)\, c(E,T),\\[3pt]
c(E,T) &= k_{\mathrm{B}}
\left( \frac{E}{k_{\mathrm{B}} T} \right)^{2}
\frac{\exp\!\left( E / k_{\mathrm{B}} T \right)}
     {\left[ \exp\!\left( E / k_{\mathrm{B}} T \right) - 1 \right]^{2}}.
\end{split}
\end{equation}
The resulting $C_{\mathrm{SW}}(T)$ curve, converted to the same unit as the measurement, is presented in Fig.~1(b) of the main text. For low temperature $k_{\mathrm{B}}T\ll\Delta_1$ [Eq.\ref{disp_eq1}], $C_{\mathrm{SW}}$ is dominated by the flat band excitations at $E=\Delta_1$ and we approximately have $C_{\mathrm{SW}}\propto(1/T^2)\exp(-\Delta_1/k_{\mathrm{B}}T)$~\cite{Xu2019_AnisotropicExchange_Nd2Zr2O7}.

\section{Determination of Interaction Parameters}\label{Sec:Determin}
Following Ref.~\cite{Xu2019_AnisotropicExchange_Nd2Zr2O7}, we extract the interaction parameters $\tilde{J}_x$, $\tilde{J}_y$, and $\tilde{J}_z$ from the analytical expressions of the spin-wave energies at the zone center $(000)$ and the zone boundary $(100)$. The two degenerate flat modes occur at the energy transfer
\begin{equation}
\Delta_1 = \sqrt{(3|\tilde{J}_z| - \tilde{J}_x)(3|\tilde{J}_z| - \tilde{J}_y)}.\label{disp_eq1}
\end{equation}
At the zone boundary $(100)$, the dispersive modes reach the energy $\Delta_2$, while at the zone center $(000)$ the higher-energy dispersive mode appears at $\Delta_3$:
\begin{align}
\Delta_2 &= \sqrt{(3|\tilde{J}_z| + \tilde{J}_x)(3|\tilde{J}_z| + \tilde{J}_y)}, \label{disp_eq2}\\
\Delta_3 &= 3\sqrt{(|\tilde{J}_z| + \tilde{J}_x)(|\tilde{J}_z| + \tilde{J}_y)}.  \label{disp_eq3}
\end{align}
By fitting energy cut taken at high symmetry points as presented in Fig.~3 (in main text) and Fig.~\ref{fig:SIFig5}, we obtain
\begin{equation}
\begin{split}
\Delta_1&=0.168(2)~\mathrm{meV}\\
\Delta_2&=0.270(4)~\mathrm{meV}\\
\Delta_3&=0.341(5)~\mathrm{meV}
\end{split}\label{Eq:levels}
\end{equation}

The uncertainties quoted in Eq.~\ref{Eq:levels} reflect statistical errors from Gaussian fits to symmetrized energy cuts. As discussed above, a small systematic uncertainty in the energy scale may arise from sample off-centering, for which the centers of the energy cuts at Bragg peaks are shifted from zero energy by up to $\sim 0.02$~meV [Fig.~\ref{fig:SIFig2}(b--d)], potentially including contributions from secondary effects. The corresponding shift at finite energy transfer is expected to be smaller and is not uniquely determined, as it depends on scattering angle and detector position. Furthermore, the use of symmetrized data over multiple symmetry-equivalent $(Q,E)$ points, which partially average such variations, complicates the reliable determination of a unique point-by-point correction or systematic uncertainty. Accordingly, this contribution is not included in the quoted uncertainties or the following error propagation. 
Including a reasonable estimate of this systematic uncertainty leads to minor increases in the parameter uncertainties and does not affect the conclusions within the present level of precision.

By solving the Eqs.\ref{disp_eq1},\ref{disp_eq2},\ref{disp_eq3}, we obtain two set of solutions,
\begin{equation}
\begin{split}
\tilde{J}_x&=0.1002^{+0.0133}_{-0.0164}~\mathrm{meV}\\
\tilde{J}_y&=-0.0012^{+0.0116}_{-0.0086}~\mathrm{meV}\\
\tilde{J}_z&=-0.0750^{+0.0017}_{-0.0016}~\mathrm{meV}
\end{split}\label{sol1}
\end{equation}
and
\begin{equation}
\begin{split}
\tilde{J}_x&=-0.0012^{+0.0116}_{-0.0086}~\mathrm{meV}\\
\tilde{J}_y&=0.1002^{+0.0133}_{-0.0164}~\mathrm{meV}\\
\tilde{J}_z&=-0.0750^{+0.0017}_{-0.0016}~\mathrm{meV}
\end{split}\label{sol2}
\end{equation}

The simulated spin-wave structure factors based on the parameters in Eq.~\ref{sol1} and Eq.~\ref{sol2} are shown in Fig.~3 and Fig.~\ref{fig:SIFig5} alongside the experimental data. After convolution with the instrumental energy resolution, the simulated linewidths exhibit excellent agreement with the measurements. Each simulated curve is rescaled by an independent overall factor (per solution) to match the total experimental intensity. Among the two models, Solution~1 (Eq.~\ref{sol1}) provides a noticeably better description of the data, particularly in reproducing the relative intensity of the lower flat modes and the higher dispersive modes. We therefore identify Solution~1 (Eq.~\ref{sol1}) as the correct set of parameters for the Nd$_2$Sn$_2$O$_7$ system. 

Using Eq.~\ref{Eqn:occ}, we numerically evaluate the reduction of the ordered moment due to zero-point quantum fluctuations in the ground state~\cite{benton2016quantum,Xu2019_AnisotropicExchange_Nd2Zr2O7}:
\begin{equation}
\begin{split}
\langle a^{\dagger}_{n,m}a_{n,m}\rangle &= 0.0180_{-0.0093}^{+0.0113}, \\
\frac{S-\langle a^{\dagger}_{n,m}a_{n,m}\rangle}{S} &= 0.9641^{+0.0187}_{-0.0226}.
\end{split}
\end{equation}
Using Eq.~\ref{Eqn:cal_m_ratio}, $m_{\mathrm{ord}}=1.708(3)~\uB/\mathrm{Nd^{3+}}$ (at $T=0.06$ K)~\cite{Bertin_2015_Nd2Sn2O7}, and $m_{\mathrm{sat}}=2.387(5)~\uB/\mathrm{Nd^{3+}}$ from our Curie-Weiss fit (see Sec.\ref{Sec:CW_detail}), we obtain the rotation-angle parameter $\vartheta$ defined in Eq.~\ref{Eqn:tau_transform}:
\begin{equation}
\vartheta = 42.3^{+1.2}_{-1.6}~^\circ=0.739_{-0.028}^{+0.021}~\mathrm{radian}.\label{Eqn:Angle}
\end{equation}
Taking together Eqs.\ref{sol1},\ref{Eqn:Angle}, we can inversely evaluate exchange parameters in the original local frame:
\begin{equation}
\begin{split}
J_x&=\tilde{J}_x\cos^2\vartheta+\tilde{J}_z\sin^2\vartheta\\
J_y&=\tilde{J}_y\\
J_z&=\tilde{J}_z\cos^2\vartheta+\tilde{J}_x\sin^2\vartheta\\
J_{xz}&=(\tilde{J}_x-\tilde{J}_z)\sin\vartheta\cos\vartheta
\end{split}
\end{equation}
which yields
\begin{equation}
\begin{split}
J_x&=0.0207_{-0.0131}^{+0.0131}~\mathrm{meV}\\
J_y&=-0.0012_{-0.0086}^{+0.0116}~\mathrm{meV}\\
J_z&=0.0045_{-0.0127}^{+0.0109}~\mathrm{meV}\\
J_{xz}&=0.0872_{-0.0095}^{+0.0077}~\mathrm{meV}
\end{split}
\end{equation}

\section{Details of Curie-Weiss fit}\label{Sec:CW_detail}
\begin{table}[tb]
\caption{Curie--Weiss fitting results for the dc susceptibility over different temperature ranges. Listed are the effective magnetic moment $\mu_{\mathrm{eff}}$ and the Curie--Weiss temperature $\theta_{\mathrm{CW}}$.}
\label{tab:CW_fits}
\centering
\begin{tabular}{ccc}
\hline\hline
Fit Range (K) & $\mu_{\mathrm{eff}}$ ($\upmu_{\mathrm{B}}$) & $\theta_{\mathrm{CW}}$ (K) \\
\hline
 2--10  & 2.397(2) & $-0.08(1)$ \\
 2--15  & 2.408(2) & $-0.13(2)$ \\
 2--20  & 2.425(3) & $-0.23(3)$ \\
 2--30  & 2.469(5) & $-0.60(7)$ \\
 2--40  & 2.514(6) & $-1.1(1)$ \\
 2--50  & 2.560(7) & $-1.7(2)$ \\
 5--15  & 2.415(3) & $-0.20(3)$ \\
 150--300 & 3.399(2) & $-61.37(32)$ \\
\hline\hline
\end{tabular}\label{Tab:CW}
\end{table}

When applying Curie--Weiss (CW) fits to the dc susceptibility with different temperature ranges, we observe variations in the extracted effective magnetic moment $\mu_{\mathrm{eff}}$ and Curie--Weiss temperature $\theta_{\mathrm{CW}}$. To obtain the effective moment of the ground-state doublet, we restrict the fitting to temperatures well below the first excited crystal-field (CEF) level of Nd$^{3+}$ in Nd$_2$Sn$_2$O$_7$, located at $\sim$26~meV~\cite{Bertin_2015_Nd2Sn2O7,Bertin_2015_thesis}. Table~\ref{Tab:CW} summarizes the fitting results for different temperature ranges. The fitted $\mu_{\mathrm{eff}}$ is largely consistent below 50~K, but gradually increases as the upper bound of the fit window is extended, reflecting the influence of higher CEF levels. By contrast, $\theta_{\mathrm{CW}}$ decreases systematically with increasing fit range.  

For quantitative analysis, we adopt the low-temperature interval $[2,10]$~K to evaluate the effective moment used in the ordered-to-saturated moment ratio [Eq.~\ref{Eqn:cal_m_ratio}]. The value of the effective moment $\mu_{\mathrm{eff}}$ extracted from the Curie--Weiss fit is then used for the saturated moment $m_{\mathrm{sat}}$. For a system in which each magnetic ion hosts a ground-state doublet with strong Ising anisotropy, characterized by a dominant $g_{\parallel}$ and negligible transverse components $g_{\perp}\!\approx\!0$, the powder-averaged effective moment is given by
\begin{equation}
\mu_{\mathrm{eff}}^2
= S(S+1)\,\frac{2g_{\perp}^2+g_{\parallel}^2}{3}\,\mu_B^2
= \frac{1}{4}g_{\parallel}^2 \mu_B^2
= m_{\mathrm{sat}}^2,
\end{equation}
with $S=\tfrac{1}{2}$~\cite{gomez2021absence}. 
We therefore take $m_{\mathrm{sat}}\!\approx\!\mu_{\mathrm{eff}}=2.397(2)\,\upmu_{\mathrm{B}}/\mathrm{Nd}^{3+}$, in good agreement with the reported $m_{\mathrm{sat}}$ values for \nzo~\cite{Xu2015_Nd2Zr2O7_CrystalField,Lhotel2015_DipoleOctupole_Nd2Zr2O7} and \nho~\cite{Anand2017_Nd2Hf2O7_muSR_INS}.

The resulting rotation angle $\vartheta$ [Eq.~\ref{Eqn:Angle}] allows us to compute the CW temperature predicted by the model~\cite{benton2016quantum,Xu2019_AnisotropicExchange_Nd2Zr2O7}:  
\begin{equation}
\begin{split}
\theta^{\mathrm{cal}}_{\mathrm{CW}} &= \frac{1}{2k_{\mathrm{B}}}\big(\tilde{J}_z \cos^2\vartheta + \tilde{J}_x \sin^2\vartheta \big) \\
&=0.026^{+0.058}_{-0.068}~\mathrm{K}
\end{split}
\end{equation}
which is roughly consistent with the fitted value $\theta_{\mathrm{CW}}=-0.08(1)$~K and with the trend that $\theta_{\mathrm{CW}}$ approaches zero when the fit is restricted to low temperature. The fitted results are also in agreement with powder measurements reported in Ref.~\cite{Bertin_2015_Nd2Sn2O7} for the ranges $[5,15]$~K and $[150,300]$~K, the latter yielding $\mu_{\mathrm{eff}}=3.399(2)~\mu_{\mathrm{B}}$, close to the free-ion value of 3.62~$\mu_{\mathrm{B}}$ for Nd$^{3+}$.

\clearpage

\bibliographystyle{apsrev4-2-titles}
\bibliography{reference}
\end{document}